\documentclass{jpsj3}
\usepackage{bm}
\usepackage{color}
\title{%
Theory of
Absorption Spectrum in the Nonequilibrium Steady State
of Superconductors
under Microwave Irradiation
}
\author{%
  Takanobu Jujo\thanks{E-mail address: jujo@ms.aist-nara.ac.jp}
}
\inst{%
  Division of Materials Science,
  Graduate School of Science and Technology,
  Nara Institute of Science and Technology,
  Ikoma, Nara 630-0101, Japan
}
\abst{%
We discuss the absorption spectrum of dirty $s$-wave superconductors
in the nonequilibrium steady state under a homogeneous and monochromatic
microwave.
In this state, there exists a finite density of states at lower
energies than the superconducting gap, and
the distribution function is also defined in this energy
region.
Because of the reduction of the distribution function
below the superconducting gap,
the absorption by thermally excited quasiparticles is
suppressed in the nonequilibrium state
compared with that in the equilibrium case.
This is in contrast to the cases of the gap function and
the absorption by direct excitation, to
which the variation of the distribution function above
the superconducting gap mainly contributes.
It is also shown that amplitude modes are generated
by frequency mixing of the pump wave and the probe wave,
and two peaks appear in the absorption spectrum. 
}
\begin{document}
\setlength{\textwidth}{504pt}
\setlength{\columnsep}{14pt}
\hoffset-23.5pt
\maketitle

  \section{Introduction}

Since an experimental study showed increases
in the critical current
of superconductors
under microwave irradiation,~\cite{dayem}
many studies have been conducted on these
nonequilibrium steady states.~\cite{mooij,dmitriev}
Theoretical calculations showed
that the enhancement of the superconducting gap can be
caused by the variation of the distribution function 
in the nonequilibrium state.~\cite{eliashberg70,ivlev}
Previously, most of the research,
including these studies, dealt with phenomena near
the superconducting critical temperature ($T_c$),
in which variations of $T_c$ and the superconducting gap are large.

Recently, phenomena at temperatures much lower than $T_c$
have attracted attention,
such as superconducting resonators containing
qubits.~\cite{wallraff,paik}
Microwaves of different frequencies 
  are used to excite the qubit and read out its state,~\cite{wallraff05}
  and it is known that there exist various modes of microwaves
  in the resonator.~\cite{rigetti}
  It has been reported that the lifetime of photons
  in the cavity is greatly increased by removing
  the two-level system,~\cite{romanenko} which is an extrinsic factor.
  It is also important to understand
  the intrinsic factor of dissipation due to
  nonequilibrium
  quasiparticle excitations in superconductors.~\cite{catelani2011}
There are many studies on the low-temperature
nonequilibrium states in which the superconducting
gap is well developed.~\cite{devisser,semenov,tikhonov}

Amplitude modes that do not couple with the electromagnetic
field within the linear response
in the equilibrium state
have also been observed by
nonlinear response techniques such as
third-harmonic generation.~\cite{matsunaga}
In nonequilibrium states,
it is expected to
be possible to observe physical quantities
that are different from those in the equilibrium state.

In this study, we investigate $s$-wave superconductors in the dirty limit
under homogeneous and monochromatic external fields
and the frequency of the pump field is set to be
smaller than the superconducting gap
with negligible direct quasiparticle excitations across
the gap.
We discuss the absorption spectrum
in this nonequilibrium steady state.
In previous studies, nonequilibrium distribution functions
were calculated
only in the range of energy above the superconducting
gap.~\cite{chang78,sridhar,gulian,goldie,devisser}
  (As for the cases other than the nonequilibrium
  state under monochromatic microwaves, formulations incorporating
  nonequilibrium corrections to the distribution function
  in all energy ranges have been used in many studies
  so far.~\cite{rainer,eschrig,jujo22})
When superconductors are irradiated with microwaves,
the density of states develops at energies below
the superconducting gap.~\cite{semenov}
In this case,
a nonequilibrium distribution function
is defined in this energy region as well.
This redistribution of quasiparticles
greatly suppresses
the absorption spectrum in the
frequency region below the absorption edge.

The nonequilibrium distribution function
is derived nonperturbatively with respect to
microwave intensity.
The frequencies of this microwave and
the probe of the absorption spectrum are different,
so collective excitation modes occur at
frequencies equivalent to
the sum and the difference of these frequencies.
Under a static magnetic field,
there is a single peak due to the amplitude
mode,~\cite{moor,nakamura,jujo22}
but in this case, it is shown that there are two peaks
in the absorption spectrum.

The structure of this paper is as follows.
In Sect. 2, using the quasiclassical approximation,
we give a formulation for describing the nonequilibrium
distribution function and the amplitude mode due to
microwave irradiation.
In Sect. 3, we give an expression for the conductivity
of the linear response
in the nonequilibrium steady state
and discuss the variation of the distribution function.
On the basis of the above formulation, in Sect. 4, we present the
results of numerical calculations
of these physical quantities.
Section 5 provides a discussion about
these calculations.

\section{Formulation}

The kinetic equations for quasiclassical Green functions
[$\hat{g}^{\pm}_{\epsilon,\epsilon'}({\mib k})$
and $\hat{g}^{(a)}_{\epsilon,\epsilon'}({\mib k})$]
are written as follows.~\cite{eliashberg}
(We set $\hbar=c=1$ in this paper with $c$ being the velocity of light.)
\begin{equation}
  \begin{split}
&  \hat{\tau}_3\epsilon\hat{g}^{\pm}_{\epsilon,\epsilon'}({\mib k})
  -\hat{g}^{\pm}_{\epsilon,\epsilon'}({\mib k})\epsilon'\hat{\tau}_3
  +\int\frac{d\omega'}{2\pi}e{\mib A}_{\omega'}\cdot{\mib v}_{\mib k}
  \left[\hat{\tau}_3\hat{g}^{\pm}_{\epsilon-\omega',\epsilon'}({\mib k})
  -\hat{g}^{\pm}_{\epsilon,\epsilon'+\omega'}({\mib k})\hat{\tau}_3\right]
\\&  -\int d\epsilon_1
  \left(\hat{\tau}_3\hat{\Sigma}^{\pm}_{\epsilon,\epsilon_1}
  \hat{g}^{\pm}_{\epsilon_1,\epsilon'}({\mib k})
  -\hat{g}^{\pm}_{\epsilon,\epsilon_1}({\mib k})
  \hat{\Sigma}^{\pm}_{\epsilon_1,\epsilon'}\hat{\tau}_3\right)=0
  \end{split}
    \label{eq:1stkineqforgpm}
\end{equation}
and
\begin{equation}
  \begin{split}
&  \hat{\tau}_3\epsilon\hat{g}^{(a)}_{\epsilon,\epsilon'}({\mib k})
  -\hat{g}^{(a)}_{\epsilon,\epsilon'}({\mib k})\epsilon'\hat{\tau}_3
  -\int d\epsilon_1
  \left(\hat{\tau}_3\hat{\Sigma}^{+}_{\epsilon,\epsilon_1}
  \hat{g}^{(a)}_{\epsilon_1,\epsilon'}({\mib k})
  -\hat{g}^{(a)}_{\epsilon,\epsilon_1}({\mib k})
  \hat{\Sigma}^{-}_{\epsilon_1,\epsilon'}\hat{\tau}_3
+\hat{\tau}_3\hat{\Sigma}^{(a)}_{\epsilon,\epsilon_1}
  \hat{g}^{-}_{\epsilon_1,\epsilon'}({\mib k})
  -\hat{g}^{+}_{\epsilon,\epsilon_1}({\mib k})
  \hat{\Sigma}^{(a)}_{\epsilon_1,\epsilon'}\hat{\tau}_3
  \right)
\\&
  +\int\frac{d\omega'}{2\pi}e{\mib A}_{\omega'}\cdot{\mib v}_{\mib k}
  \left[\hat{\tau}_3\hat{g}^{(a)}_{\epsilon-\omega',\epsilon'}({\mib k})
    +\hat{\tau}_3
    (T^h_{\epsilon}-T^h_{\epsilon-\omega'})
    \hat{g}^{-}_{\epsilon-\omega',\epsilon'}({\mib k})
    -\hat{g}^{(a)}_{\epsilon,\epsilon'+\omega'}({\mib k})\hat{\tau}_3
    -(T^h_{\epsilon'+\omega'}-T^h_{\epsilon'})
    \hat{g}^{+}_{\epsilon,\epsilon'+\omega'}({\mib k})\hat{\tau}_3
    \right]
  =0.
  \end{split}
  \label{eq:1stkineqforga}
\end{equation}
The superscripts $+$, $-$, and $(a)$ mean retarded,
advanced, and anomalous terms, respectively,
and 
$\hat{\cdot}$ indicates a $2\times 2$ matrix in the Nambu space.
$\hat{\tau}_3=
\left(\begin{smallmatrix} 1 & 0\\0 & -1\end{smallmatrix}\right)$
  and
$T^h_{\epsilon}:={\rm tanh}(\epsilon/2T)$,
  $T$ being temperature.
  ${\mib v}_{\mib k}$ and ${\mib A}_{\omega}$
  are the quasiparticle velocity
  and the external field, respectively.
Here, we take account of the electron--phonon
interaction with the weak-coupling approximation,
the scattering by nonmagnetic impurities
with the t-matrix approximation,
and the effect of inelastic scatterings
($\hat{\Sigma}^{inel}$), which are specified below.
The self-energies are written as 
\begin{equation}
  \hat{\Sigma}^{\pm}_{\epsilon,\epsilon'}:=
  p\hat{\Sigma}^{p}_{\epsilon-\epsilon'}
  +\frac{1}{2\tau}\int_{FS}\hat{\tau}_3\hat{g}^{\pm}_{\epsilon,\epsilon'}({\mib k})
  \hat{\tau}_3
  +\hat{\Sigma}^{inel,\pm}_{\epsilon,\epsilon'}
  \label{eq:sfekpm}
\end{equation}
and
\begin{equation}
  \hat{\Sigma}^{(a)}_{\epsilon,\epsilon'}:=
  (T^h_{\epsilon}-T^h_{\epsilon'})p\hat{\Sigma}^p_{\epsilon-\epsilon'}
  +\frac{1}{2\tau}
  \int_{FS}\hat{\tau}_3\hat{g}^{(a)}_{\epsilon,\epsilon'}({\mib k})
  \hat{\tau}_3
  +\hat{\Sigma}^{inel(a)}_{\epsilon,\epsilon'}
  \label{eq:sfeka}
\end{equation}
with $p:=(mk_F/2\pi)(g_{ph}^2/\omega_D)$ 
($g_{ph}$ is
the electron--phonon coupling constant 
and $\omega_D$ is the Debye frequency) describing
the attractive interaction between electrons,
$1/2\tau:=(mk_F/2\pi)n_i u_i^2
  /[1+(u_i mk_F/2\pi)^2]$
($n_i$ is the concentration and $u_i$ is the potential
of nonmagnetic impurities) meaning the elastic scattering rate,
and $\int_{FS}$ the integration over the Fermi surface
($m$ and $k_F$ are the quasiparticle mass and
the Fermi wave number, respectively).
Here,
\begin{equation}
  \hat{\Sigma}^p_{\omega}:=
  \int\frac{d\epsilon}{2\pi i}\int_{FS}\hat{\tau}_3\left[
    T^h_{\epsilon-\omega/2}
    \hat{g}^+_{\epsilon+\omega/2,\epsilon-\omega/2}({\mib k})
    -T^h_{\epsilon+\omega/2}
    \hat{g}^-_{\epsilon+\omega/2,\epsilon-\omega/2}({\mib k})
    +\hat{g}^{(a)}_{\epsilon+\omega/2,\epsilon-\omega/2}({\mib k})
    \right]\hat{\tau}_3
  \label{eq:eqforsigmap}
\end{equation}
($i=\sqrt{-1}$).

Following the derivation of the Usadel equation
in the dirty limit ($\Delta\tau\ll 1$)~\cite{usadel},
we substitute 
$\hat{g}^{\pm}_{\epsilon,\epsilon'}({\mib k})
=\hat{g}^{\pm}_{\epsilon,\epsilon'}
+\sum_{\mu}v^{\mu}_{\mib k}\hat{g}^{\pm'\mu}_{\epsilon,\epsilon'}$
and
$\hat{g}^{(a)}_{\epsilon,\epsilon'}({\mib k})
=\hat{g}^{(a)}_{\epsilon,\epsilon'}
+\sum_{\mu}v^{\mu}_{\mib k}\hat{g}^{(a)'\mu}_{\epsilon,\epsilon'}$
into Eqs. (\ref{eq:1stkineqforgpm}) and (\ref{eq:1stkineqforga})
and use the following normalization relations.
[$v^{\mu}_{\mib k}$ is one of the components of
  ${\mib v}_{\mib k}=(v_{\mib k}^x,v_{\mib k}^y,v_{\mib k}^z)$.
    The superscripts ${\mu}$ in 
    $\hat{g}^{\pm'\mu}_{\epsilon,\epsilon'}$
    and $\hat{g}^{(a)'\mu}_{\epsilon,\epsilon'}$
    are omitted hereafter.]
\begin{equation}
  \int d\epsilon_1\hat{g}^{\pm}_{\epsilon,\epsilon_1}
  \hat{\tau}_3\hat{g}^{\pm}_{\epsilon_1,\epsilon'}
  =-\hat{\tau}_3\delta(\epsilon-\epsilon'),
  \label{eq:normalizationpm}
\end{equation}
\begin{equation}
  \int d\epsilon_1\left(\hat{g}^{+}_{\epsilon,\epsilon_1}
  \hat{\tau}_3\hat{g}^{(a)}_{\epsilon_1,\epsilon'}
  +\hat{g}^{(a)}_{\epsilon,\epsilon_1}
  \hat{\tau}_3\hat{g}^{-}_{\epsilon_1,\epsilon'}\right)
  =0,
    \label{eq:normalizationa}
\end{equation}
\begin{equation}
  \int d\epsilon_1\left(\hat{g}^{\pm}_{\epsilon,\epsilon_1}
  \hat{\tau}_3\hat{g}^{\pm'}_{\epsilon_1,\epsilon'}
  +\hat{g}^{\pm'}_{\epsilon,\epsilon_1}
  \hat{\tau}_3\hat{g}^{\pm}_{\epsilon_1,\epsilon'}\right)
  =0,
\end{equation}
and
\begin{equation}
  \int d\epsilon_1\left(
  \hat{g}^{+}_{\epsilon,\epsilon_1}
  \hat{\tau}_3\hat{g}^{(a)'}_{\epsilon_1,\epsilon'}
  +\hat{g}^{(a)'}_{\epsilon,\epsilon_1}
  \hat{\tau}_3\hat{g}^{-}_{\epsilon_1,\epsilon'}
  +
  \hat{g}^{+'}_{\epsilon,\epsilon_1}
  \hat{\tau}_3\hat{g}^{(a)}_{\epsilon_1,\epsilon'}
  +\hat{g}^{(a)}_{\epsilon,\epsilon_1}
  \hat{\tau}_3\hat{g}^{-'}_{\epsilon_1,\epsilon'}
  \right)
  =0
\end{equation}
[$\delta(\cdot)$ is the delta function].
Then the kinetic equations for quasiclassical Green functions
[Eqs. (\ref{eq:1stkineqforgpm}) and (\ref{eq:1stkineqforga})]
are given by the following equations
($A^{\mu}_{\omega}$ is written as $A_{\omega}$):
\begin{equation}
  \begin{split}
  \hat{g}^{\pm'}_{\epsilon,\epsilon'}=
  -\tau\int\frac{d\omega'}{2\pi}
  eA_{\omega'}
  \Bigl\{
  \delta(\epsilon-\epsilon'-\omega')
  +\int d\epsilon_1
    \hat{g}^{\pm}_{\epsilon,\epsilon_1}
    \hat{g}^{\pm}_{\epsilon_1-\omega',\epsilon'}
  \Bigr\},
  \end{split}
  \label{eq:approxgpmd}
\end{equation}
\begin{equation}
  \begin{split}
&  \hat{g}^{(a)'}_{\epsilon,\epsilon'}=
  -\tau\int\frac{d\omega'}{2\pi}
  eA_{\omega'}
  \Bigl\{
  (T^h_{\epsilon}-T^h_{\epsilon'})
  \delta(\epsilon-\epsilon'-\omega')
\\&  +\int d\epsilon_1
  \left[
    \hat{g}^+_{\epsilon,\epsilon_1}
    \hat{g}^{(a)}_{\epsilon_1-\omega',\epsilon'}
    +
        \hat{g}^{(a)}_{\epsilon,\epsilon_1}
        \hat{g}^{-}_{\epsilon_1-\omega',\epsilon'}
        +
        (T^h_{\epsilon_1}-T^h_{\epsilon_1-\omega'})
        \hat{g}^{+}_{\epsilon,\epsilon_1}
        \hat{g}^{-}_{\epsilon_1-\omega',\epsilon'}
        \right]
  \Bigr\},
  \end{split}
  \label{eq:approxgad}
\end{equation}
  \begin{equation}
  \begin{split}
    &  \hat{\tau}_3
    \epsilon
    \hat{g}^{\pm}_{\epsilon,\epsilon'}
    -\hat{g}^{\pm}_{\epsilon,\epsilon'}
        \epsilon'
    \hat{\tau}_3
  -\int d\epsilon_1
  \left(
  \hat{\tau}_3\hat{\Sigma}^{ep,\pm}_{\epsilon,\epsilon_1}
  \hat{g}^{\pm}_{\epsilon_1,\epsilon'}
  -\hat{g}^{\pm}_{\epsilon,\epsilon_1}
  \hat{\Sigma}^{ep,\pm}_{\epsilon_1,\epsilon'}\hat{\tau}_3
  \right)
  \\&  -e^2D
  \int\frac{d\omega_1 d\omega_2}{(2\pi)^2}
  A_{\omega_1}A_{\omega_2}\int d\epsilon_1
\left(
  \hat{\tau}_3
    \hat{g}^{\pm}_{\epsilon-\omega_1,\epsilon_1}
    \hat{g}^{\pm}_{\epsilon_1-\omega_2,\epsilon'}
  -
        \hat{g}^{\pm}_{\epsilon,\epsilon_1}
        \hat{g}^{\pm}_{\epsilon_1-\omega_2,\epsilon'+\omega_1}
 \hat{\tau}_3
\right)
  =0,
  \end{split}
  \label{eq:approxgpm}
  \end{equation}
  and
\begin{equation}
  \begin{split}
    &  \hat{\tau}_3    \epsilon
    \hat{g}^{(a)}_{\epsilon,\epsilon'}
    -\hat{g}^{(a)}_{\epsilon,\epsilon'}
    \epsilon'    \hat{\tau}_3
-e^2D
  \int\frac{d\omega_1 d\omega_2}{(2\pi)^2}
  A_{\omega_1}A_{\omega_2}\int d\epsilon_1
  \Bigl\{
\\&  \hat{\tau}_3
  \left[
    \hat{g}^{+}_{\epsilon-\omega_1,\epsilon_1}
    \hat{g}^{(a)}_{\epsilon_1-\omega_2,\epsilon'}
    +    \hat{g}^{(a)}_{\epsilon-\omega_1,\epsilon_1}
  \hat{g}^{-}_{\epsilon_1-\omega_2,\epsilon'}
+    (T^h_{\epsilon_1}-T^h_{\epsilon_1-\omega_2})
    \hat{g}^{+}_{\epsilon-\omega_1,\epsilon_1}
    \hat{g}^{-}_{\epsilon_1-\omega_2,\epsilon'}
    +    (T^h_{\epsilon}-T^h_{\epsilon-\omega_1})
    \hat{g}^{-}_{\epsilon-\omega_1,\epsilon_1}
    \hat{g}^{-}_{\epsilon_1-\omega_2,\epsilon'}
  \right]
\\&  -\left[
        \hat{g}^{+}_{\epsilon,\epsilon_1}
        \hat{g}^{(a)}_{\epsilon_1-\omega_2,\epsilon'+\omega_1}
        +        \hat{g}^{(a)}_{\epsilon,\epsilon_1}
    \hat{g}^{-}_{\epsilon_1-\omega_2,\epsilon'+\omega_1}
+    (T^h_{\epsilon_1}-T^h_{\epsilon_1-\omega_2})
    \hat{g}^{+}_{\epsilon,\epsilon_1}
    \hat{g}^{-}_{\epsilon_1-\omega_2,\epsilon'+\omega_1}
    +    (T^h_{\epsilon'+\omega_1}-T^h_{\epsilon'})
    \hat{g}^{+}_{\epsilon,\epsilon_1}
    \hat{g}^{+}_{\epsilon_1-\omega_2,\epsilon'+\omega_1}
  \right]\hat{\tau}_3
  \Bigr\}
\\&  -\int d\epsilon_1
  \left(
  \hat{\tau}_3\hat{\Sigma}^{ep,+}_{\epsilon,\epsilon_1}
  \hat{g}^{(a)}_{\epsilon_1,\epsilon'}
  -\hat{g}^{(a)}_{\epsilon,\epsilon_1}
  \hat{\Sigma}^{ep,-}_{\epsilon_1,\epsilon'}\hat{\tau}_3
  +
    \hat{\tau}_3\hat{\Sigma}^{ep(a)}_{\epsilon,\epsilon_1}
  \hat{g}^{-}_{\epsilon_1,\epsilon'}
  -\hat{g}^{+}_{\epsilon,\epsilon_1}
  \hat{\Sigma}^{ep(a)}_{\epsilon_1,\epsilon'}\hat{\tau}_3
  \right)
  =0.
  \end{split}
  \label{eq:approxga}
\end{equation}
$D:=v_F^2\tau/3$ is the diffusion constant,
\begin{equation}
  \hat{\Sigma}^{ep,\pm}_{\epsilon,\epsilon'}:=
  p\hat{\Sigma}^{p}_{\epsilon-\epsilon'}
  +\hat{\Sigma}^{inel,\pm}_{\epsilon,\epsilon'},
  \label{eq:sfepm}
\end{equation}
and 
\begin{equation}
    \hat{\Sigma}^{ep(a)}_{\epsilon,\epsilon'}:=
  (T^h_{\epsilon}-T^h_{\epsilon'})p\hat{\Sigma}^p_{\epsilon-\epsilon'}
    +\hat{\Sigma}^{inel(a)}_{\epsilon,\epsilon'}
  \label{eq:sfea}
\end{equation}
from Eqs. (\ref{eq:sfekpm}) and (\ref{eq:sfeka}).

Here, we introduce monochromatic pump ($A_{\pm\omega_0}$)
and probe ($A^{pr}_{\omega}$) fields as 
\begin{equation}
  A_{\omega'}=2\pi
  \left[
    \sum_{\pm}A_{\pm\omega_0}\delta(\omega'\mp\omega_0)
  +A^{pr}_{\omega}\delta(\omega'-\omega)
  \right]
\end{equation}
and substitute this $A_{\omega}$ into
the above kinetic equations.
(The probe field is set to be parallel to
the pump field in this work.)
We assume that $e^2DA_{\omega_0}A_{-\omega_0}\ll \omega_0$
[under this condition, the term
that is proportional to $\delta(\epsilon-\epsilon'\pm n\omega_0)$
with $n\ge 2$ can be neglected~\cite{semenov}].
We perform 
the perturbative expansion by the probe field
and take account of
the zeroth and first-order terms
of the probe field (the linear response).
Then quasiclassical Green functions can be
written as 
\begin{equation}
  \hat{g}^{\pm}_{\epsilon,\epsilon'}\simeq
\hat{g}^{\pm}_{\epsilon}\delta(\epsilon-\epsilon')
+\sum_{s=\pm 1}
\hat{g}^{s,\pm}_{\epsilon,\epsilon'}\delta(\epsilon-\epsilon'-\omega_{s}),
\label{eq:gpmquasi}
\end{equation}
\begin{equation}
  \hat{g}^{(a)}_{\epsilon,\epsilon'}\simeq
\hat{g}^{(a)}_{\epsilon}\delta(\epsilon-\epsilon')
+\sum_{s=\pm 1}
\hat{g}^{s(a)}_{\epsilon,\epsilon'}\delta(\epsilon-\epsilon'-\omega_{s}),
\label{eq:gaquasi}
\end{equation}
\begin{equation}
  \hat{\Sigma}^{ep,\pm}_{\epsilon,\epsilon'}\simeq
  \left(\hat{\Sigma}^{\gamma,\pm}_{\epsilon}
+\Delta\hat{\tau}_1\right)
  \delta(\epsilon-\epsilon')
+\sum_{s=\pm 1}
\hat{\Sigma}^{s,\pm}_{\epsilon,\epsilon'}\delta(\epsilon-\epsilon'-\omega_{s}),
\label{eq:sigeppm}
\end{equation}
and
\begin{equation}
  \hat{\Sigma}^{ep(a)}_{\epsilon,\epsilon'}\simeq
\hat{\Sigma}^{\gamma(a)}_{\epsilon}\delta(\epsilon-\epsilon')
+\sum_{s=\pm 1}
\hat{\Sigma}^{s(a)}_{\epsilon,\epsilon'}\delta(\epsilon-\epsilon'-\omega_{s}).
\label{eq:sigepa}
\end{equation}
[In the right-hand sides of Eqs. (\ref{eq:gpmquasi})--(\ref{eq:sigepa}),
the second terms are linear in the probe field and 
the first terms do not contain the probe field.
The pump field is taken into account
nonperturbatively in both terms, which
causes the system to be in the nonequilibrium steady state.]
Here,\begin{equation}
  \omega_{\pm 1}:=\omega\pm\omega_0
  \label{eq:defoms}
\end{equation}
(in the case of $\omega=\omega_0$ where $\omega_{-1}=0$,
the probe wave cannot be distinguished from the pump wave,
and this point is discussed in Sect. 4),
and $\Delta$ is the superconducting gap determined by
the equation obtained from Eq. (\ref{eq:eqforsigmap}),
\begin{equation}
  \Delta\hat{\tau}_1=
  p  \int\frac{d\epsilon}{2\pi i}
  \hat{\tau}_3\left[
  T^h_{\epsilon}
  \left(\hat{g}^+_{\epsilon}-\hat{g}^-_{\epsilon}\right)
    +\hat{g}^{(a)}_{\epsilon}
    \right]\hat{\tau}_3,
  \label{eq:tau1gapeq}
\end{equation}
with $\tau_1=\left(\begin{smallmatrix}0&1\\1&0\end{smallmatrix}\right)$.
By substituting Eqs. (\ref{eq:gpmquasi})--(\ref{eq:sigepa})
into Eqs. (\ref{eq:approxgpm}) and (\ref{eq:approxga}),
we obtain the following equations:
\begin{equation}
        \hat{\tau}_3\left(
        \tilde{\epsilon}^{\pm}\hat{\tau}_0
        -\tilde{\Delta}^{\pm}_{\epsilon}\hat{\tau}_1
    \right)\hat{g}^{\pm}_{\epsilon}
    -\hat{g}^{\pm}_{\epsilon}
    \left(\tilde{\epsilon}^{\pm}\hat{\tau}_0
    -\tilde{\Delta}^{\pm}_{\epsilon}{\tau}_1
    \right)\hat{\tau}_3=0,
        \label{eq:4thkineqforg0pm}
\end{equation}
\begin{equation}
  \begin{split}
    &  \hat{\tau}_3\left(
    \tilde{\epsilon}^+\hat{\tau}_0
    -\tilde{\Delta}^+_{\epsilon}\hat{\tau}_1
    \right)\hat{g}^{(a)}_{\epsilon}
    -\hat{g}^{(a)}_{\epsilon}
    \left(\tilde{\epsilon}^{-}\hat{\tau}_0
    -\tilde{\Delta}^-_{\epsilon}{\tau}_1
    \right)\hat{\tau}_3
  +
    \hat{\tau}_3\hat{Z}^{(a)}_{\epsilon}
  \hat{g}^{-}_{\epsilon}
  -\hat{g}^{+}_{\epsilon}
  \hat{Z}^{(a)}_{\epsilon}\hat{\tau}_3
=0,
  \end{split}
      \label{eq:4thkineqforga0}
\end{equation}
\begin{equation}
  \begin{split}
    &  \hat{\tau}_3\left(
    \tilde{\epsilon_s}^{\pm}\hat{\tau}_0
    -\tilde{\Delta}^{\pm}_{\epsilon_s}\hat{\tau}_1
    \right)\hat{g}^{s,\pm}_{\epsilon_s,\epsilon'_s}
    -\hat{g}^{s,\pm}_{\epsilon_s,\epsilon'_s}
    \left(\tilde{\epsilon'_s}^{\pm}\hat{\tau}_0
    -\tilde{\Delta}^{\pm}_{\epsilon_s'}{\tau}_1
    \right)\hat{\tau}_3
+
    \hat{\tau}_3\hat{Y}^{s,\pm}_{\epsilon_s,\epsilon'_s}
  \hat{g}^{\pm}_{\epsilon'_s}
  -\hat{g}^{\pm}_{\epsilon_s}
  \hat{Y}^{s,\pm}_{\epsilon_s,\epsilon'_s}\hat{\tau}_3
  =0,
  \end{split}
      \label{eq:4thkineqforgppm}
\end{equation}
and
\begin{equation}
  \begin{split}
    &  \hat{\tau}_3\left(
    \tilde{\epsilon_s}^+\hat{\tau}_0
    -\tilde{\Delta}^+_{\epsilon_s}\hat{\tau}_1
    \right)\hat{g}^{s(a)}_{\epsilon_s,\epsilon'_s}
    -\hat{g}^{s(a)}_{\epsilon_s,\epsilon'_s}
    \left(\tilde{\epsilon'_s}^{-}\hat{\tau}_0
    -\tilde{\Delta}^-_{\epsilon_s'}{\tau}_1
    \right)\hat{\tau}_3
+    \hat{\tau}_3\hat{Z}^{(a)}_{\epsilon_s}
  \hat{g}^{s,-}_{\epsilon_s,\epsilon'_s}
  -\hat{g}^{s,+}_{\epsilon_s,\epsilon'_s}
  \hat{Z}^{(a)}_{\epsilon'_s}\hat{\tau}_3
  \\&+
    \hat{\tau}_3\hat{Y}^{s,+}_{\epsilon_s,\epsilon'_s}
  \hat{g}^{(a)}_{\epsilon'_s}
  -\hat{g}^{(a)}_{\epsilon_s}
  \hat{Y}^{s,-}_{\epsilon_s,\epsilon'_s}\hat{\tau}_3
+
    \hat{\tau}_3\hat{X}^{s(a)}_{\epsilon_s,\epsilon'_s}
  \hat{g}^{-}_{\epsilon'_s}
  -\hat{g}^{+}_{\epsilon_s}
  \hat{X}^{s(a)}_{\epsilon_s,\epsilon'_s}\hat{\tau}_3
  =0.
  \end{split}
    \label{eq:4thkineqforgpa}
\end{equation}
Here,
\begin{equation}
\epsilon_{s}=\epsilon+\omega_{s}/2,\quad
\epsilon'_{s}=\epsilon-\omega_{s}/2
\quad (s=\pm 1),
\label{eq:defepssd}
\end{equation}
\begin{equation}
\tilde{\epsilon}^{\pm}=\epsilon
-\Sigma_{n,\epsilon}^{\pm}
-\alpha
(g^{\pm}_{\epsilon+\omega_0}+g^{\pm}_{\epsilon-\omega_0}),
\label{eq:deftileps}
\end{equation}
and
\begin{equation}
\tilde{\Delta}^{\pm}_{\epsilon}=\Delta
+\Sigma_{a,\epsilon}^{\pm}
+\alpha
(f^{\pm}_{\epsilon+\omega_0}+f^{\pm}_{\epsilon-\omega_0})
\label{eq:deftildel}
\end{equation}
with
\begin{equation}
  \alpha:=e^2DA_{\omega_0}A_{-\omega_0},
  \label{eq:defalpha}
  \end{equation}
$\hat{\Sigma}^{\gamma,\pm}_{\epsilon}:=
\Sigma^{\pm}_{n,\epsilon}\hat{\tau}_0
+\Sigma^{\pm}_{a,\epsilon}\hat{\tau}_1$
is the inelastic scattering effect,
$\hat{g}^{\pm}_{\epsilon}=
g^{\pm}_{\epsilon}\hat{\tau}_0+f^{\pm}_{\epsilon}\hat{\tau}_1$
[$\hat{\tau}_0=\left(\begin{smallmatrix}1&0\\0&1\end{smallmatrix}\right)$,
$g^{\pm}_{\epsilon}=-\tilde{\epsilon}^{\pm}/\sqrt{
(\tilde{\Delta}^{\pm}_{\epsilon})^2
-(\tilde{\epsilon}^{\pm})^2}$,
and
$f^{\pm}_{\epsilon}=-\tilde{\Delta}^{\pm}_{\epsilon}/\sqrt{
(\tilde{\Delta}^{\pm}_{\epsilon})^2
-(\tilde{\epsilon}^{\pm})^2}$],
\begin{equation}
  \begin{split}
    &  \hat{X}^{s(a)}_{\epsilon_s,\epsilon'_s}
    =-\hat{\Sigma}^{s(a)}_{\epsilon_s,\epsilon'_s}
  -e^2DA_{s\omega_0}A^{pr}_{\omega}
  \left[
(\tilde{T}^h_{\epsilon_{-s}}-T^h_{\epsilon'_s})\hat{g}^+_{\epsilon_{-s}}
+(T^h_{\epsilon_s}-\tilde{T}^h_{\epsilon_{-s}})\hat{g}^-_{\epsilon_{-s}}
+(\tilde{T}^h_{\epsilon'_{-s}}-T^h_{\epsilon'_s})\hat{g}^+_{\epsilon'_{-s}}
+(T^h_{\epsilon_s}-\tilde{T}^h_{\epsilon'_{-s}})\hat{g}^-_{\epsilon'_{-s}}
    \right]
\\&    -\alpha
\sum_{\pm}\left[
       \hat{g}^{s(a)}_{\epsilon_s\pm\omega_0,\epsilon'_s\pm\omega_0}
    +    (T^h_{\epsilon'_s\pm\omega_0}-T^h_{\epsilon'_s})
  \hat{g}^{s,+}_{\epsilon_s\pm\omega_0,\epsilon'_s\pm\omega_0}
   +(T^h_{\epsilon_s}-T^h_{\epsilon_s\pm\omega_0})
    \hat{g}^{s,-}_{\epsilon_s\pm\omega_0,\epsilon'_s\pm\omega_0}
     \right],
  \end{split}
  \label{eq:defhatx}
\end{equation}
\begin{equation}
  \hat{Y}^{s,\pm}_{\epsilon_s,\epsilon'_s}
  =-\hat{\Sigma}^{s,\pm}_{\epsilon_s,\epsilon'_s}
  -e^2DA_{s\omega_0}A^{pr}_{\omega}
  \left[
      \hat{g}^{\pm}_{\epsilon_{-s}}+\hat{g}^{\pm}_{\epsilon'_{-s}}
      \right]
  -\alpha
  \left[
    \hat{g}^{s,\pm}_{\epsilon_s-\omega_0,\epsilon'_s-\omega_0}
    +\hat{g}^{s,\pm}_{\epsilon_s+\omega_0,\epsilon'_s+\omega_0}
    \right],
    \label{eq:defhaty}
\end{equation}
and
\begin{equation}
  \hat{Z}^{(a)}_{\epsilon}=-\hat{\Sigma}^{\gamma(a)}_{\epsilon}
  -\alpha
\sum_{\pm}
\left(\tilde{T}^h_{\epsilon\pm\omega_0}
-T^h_{\epsilon}\right)
      \left(\hat{g}^+_{\epsilon\pm\omega_0}-\hat{g}^-_{\epsilon\pm\omega_0}
      \right).
\end{equation}
Here,
\begin{equation}
  \tilde{T}^h_{\epsilon}:=T^h_{\epsilon}+x^{(a)}_{\epsilon},
  \label{eq:deftilth}
\end{equation}
where $x^{(a)}_{\epsilon}$ means
the nonequilibrium correction term to the electronic distribution function
and is defined as 
\begin{equation}
x^{(a)}_{\epsilon}:=  \frac{g^{(a)}_{\epsilon}}{g^+_{\epsilon}-g^-_{\epsilon}}
  =\frac{f^{(a)}_{\epsilon}}{f^+_{\epsilon}-f^-_{\epsilon}}
  \label{eq:definitionxa}
\end{equation}
(the normalization condition is used here).
With the use of Eq. (\ref{eq:definitionxa}),
the kinetic equation 
(\ref{eq:4thkineqforga0})
is written as 
\begin{equation}
  \begin{split}
    & 
    x^{(a)}_{\epsilon}
    =
    \frac{-1}
        {\left(\tilde{\epsilon}^+-\tilde{\epsilon}^-\right)
\left(g^+_{\epsilon}-g^-_{\epsilon}\right)
    -(\tilde{\Delta}^+_{\epsilon} -\tilde{\Delta}^-_{\epsilon})
    \left(f^+_{\epsilon}-f^-_{\epsilon}\right)}
        \Bigl\{
        (g^+_{\epsilon}-g^-_{\epsilon})\Sigma^{(a)}_{n,\epsilon}
        +(f^+_{\epsilon}-f^-_{\epsilon})\Sigma^{(a)}_{a,\epsilon}
        \\&
        +
        \alpha
        \sum_{\pm}\left\{
(\tilde{T}^h_{\epsilon\pm\omega_0}-T^h_{\epsilon})
\left[
(g^{+}_{\epsilon\pm\omega_0}-g^{-}_{\epsilon\pm\omega_0})
  (g^+_{\epsilon}-g^-_{\epsilon})
  +
  (f^{+}_{\epsilon\pm\omega_0}-f^{-}_{\epsilon\pm\omega_0})
  (f^+_{\epsilon}-f^-_{\epsilon})
  \right]
\right\}
\Bigr\}
  \end{split}
  \label{eq:eqforxa}
\end{equation}
with $\hat{\Sigma}^{\gamma(a)}_{\epsilon}=
\Sigma^{(a)}_{n,\epsilon}\hat{\tau}_0
+\Sigma^{(a)}_{a,\epsilon}\hat{\tau}_1$.

\subsection{Vertex correction}

The vertex corrections in Eqs. (\ref{eq:defhatx})
and (\ref{eq:defhaty}) are obtained
from Eq. (\ref{eq:eqforsigmap}) as 
\begin{equation}
  \hat{\Sigma}^{s,+}_{\epsilon_s,\epsilon_s'}
  =\hat{\Sigma}^{s,-}_{\epsilon_s,\epsilon_s'}
  =p\int_{-\infty}^{\infty}\frac{d\epsilon}{2\pi i}
  \hat{\tau}_3
\left[
T^h_{\epsilon_s'}\hat{g}^{s,+}_{\epsilon_s,\epsilon'_s}
-T^h_{\epsilon_s}\hat{g}^{s,-}_{\epsilon_s,\epsilon'_s}
+\hat{g}^{s(a)}_{\epsilon_s,\epsilon'_s}\right]
  \hat{\tau}_3
  =:\hat{\Sigma}^{s}_{\omega_s}
\end{equation}
and
\begin{equation}
\hat{\Sigma}^{s(a)}_{\epsilon_s,\epsilon_s'}=
(T^h_{\epsilon_s}-T^h_{\epsilon'_s})
\hat{\Sigma}^{s}_{\omega_s}.
\end{equation}
The components of the matrix 
($\hat{\Sigma}^s_{\omega_s}=
\Sigma^s_{n,\omega_s}\hat{\tau}_0
+\Sigma^s_{a,\omega_s}\hat{\tau}_1$)
are given by
\begin{equation}
  \begin{pmatrix}
    \Sigma_{n,\omega_s}^{s}\\\Sigma_{a,\omega_s}^{s}
  \end{pmatrix}
  =
  p\int_{-\infty}^{\infty}\frac{d\epsilon}{2\pi i}
  \hat{\tau}_3\left[
    \begin{pmatrix}
      g^{s(a)}_{\epsilon_s,\epsilon_s'}
      \\f^{s(a)}_{\epsilon_s,\epsilon_s'}
    \end{pmatrix}
    +
    T^h_{\epsilon_s'}
        \begin{pmatrix}
      g^{s,+}_{\epsilon_s,\epsilon_s'}
      \\f^{s,+}_{\epsilon_s,\epsilon_s'}
        \end{pmatrix}
        -
        T^h_{\epsilon_s}
        \begin{pmatrix}
      g^{s,-}_{\epsilon_s,\epsilon_s'}
      \\f^{s,-}_{\epsilon_s,\epsilon_s'}
        \end{pmatrix}
        \right]
\end{equation}
with $\hat{g}^{s,\pm}_{\epsilon,\epsilon'}=
g^{s,\pm}_{\epsilon,\epsilon'}\hat{\tau}_0
+f^{s,\pm}_{\epsilon,\epsilon'}\hat{\tau}_1$
and
$\hat{g}^{s(a)}_{\epsilon,\epsilon'}=
g^{s(a)}_{\epsilon,\epsilon'}\hat{\tau}_0
+f^{s(a)}_{\epsilon,\epsilon'}\hat{\tau}_1$.
By introducing the following quantities with
the use of normalization conditions, 
\begin{equation}
y^{s,\pm}_{\epsilon,\epsilon'}:=  \frac{g^{s,\pm}_{\epsilon,\epsilon'}}{g^{\pm}_{\epsilon}-g^{\pm}_{\epsilon'}}
 =\frac{f^{s,\pm}_{\epsilon,\epsilon'}}{f^{\pm}_{\epsilon}-f^{\pm}_{\epsilon'}}
   \label{eq:definitionyp}
\end{equation}
and
\begin{equation}
z^{s(a)}_{\epsilon,\epsilon'}:=  \frac{g^{s(a)}_{\epsilon,\epsilon'}
    -x^{(a)}_{\epsilon'}y^{s,+}_{\epsilon,\epsilon'}
     (g^+_{\epsilon}-g^+_{\epsilon'}) 
    +x^{(a)}_{\epsilon}y^{s,-}_{\epsilon,\epsilon'}
    (g^-_{\epsilon}-g^-_{\epsilon'})}{g^+_{\epsilon}-g^-_{\epsilon'}}
  =
    \frac{f^{s(a)}_{\epsilon,\epsilon'}
    -x^{(a)}_{\epsilon'}y^{s,+}_{\epsilon,\epsilon'}
     (f^+_{\epsilon}-f^+_{\epsilon'}) 
    +x^{(a)}_{\epsilon}y^{s,-}_{\epsilon,\epsilon'}
    (f^-_{\epsilon}-f^-_{\epsilon'})}{f^+_{\epsilon}-f^-_{\epsilon'}}
      \label{eq:definitionzp}
\end{equation}
($s=\pm 1$),
the kinetic equations 
(\ref{eq:4thkineqforgppm}) and  
  (\ref{eq:4thkineqforgpa})
are rewritten as follows:
\begin{equation}
  \begin{split}
    &
    \begin{pmatrix}
      g^{\pm}_{\epsilon_s}-g^{\pm}_{\epsilon_s'}
      \\
      f^{\pm}_{\epsilon_s}-f^{\pm}_{\epsilon_s'}
    \end{pmatrix}
    y^{s,\pm}_{\epsilon_s,\epsilon_s'}
         =
\hat{M}^{\pm,\pm}_{\epsilon_s,\epsilon'_s}
      \Biggl[
      \begin{pmatrix}
       \Sigma^{s}_{n,\omega_s}
 \\
 \Sigma^{s}_{a,\omega_s}
 \end{pmatrix}
              +e^2DA_{s\omega_0}A^{pr}_{\omega}
                \begin{pmatrix}
  g^{\pm}_{\epsilon_{-s}}+g^{\pm}_{\epsilon'_{-s}}
\\
f^{\pm}_{\epsilon_{-s}}+f^{\pm}_{\epsilon'_{-s}}
              \end{pmatrix}
                \\&      +\alpha
                \sum_{t=\pm 1}
\begin{pmatrix}
  g^{\pm}_{\epsilon_s+t\omega_0}
  -g^{\pm}_{\epsilon'_s+t\omega_0}
  \\f^{\pm}_{\epsilon_s+t\omega_0}
  -f^{\pm}_{\epsilon'_s+t\omega_0}
\end{pmatrix}
y^{s,\pm}_{\epsilon_s+t\omega_0,\epsilon'_s+t\omega_0}
\Biggr]
  \end{split}
  \label{eq:eqforyspm}
\end{equation}
and
\begin{equation}
  \begin{split}
    &\begin{pmatrix}
         g_{\epsilon_s}^+ -g_{\epsilon_s'}^-\\
         f_{\epsilon_s}^+ -f_{\epsilon_s'}^-\end{pmatrix}
    z^{s(a)}_{\epsilon_s,\epsilon_s'}=
\hat{M}^{+,-}_{\epsilon_s,\epsilon_s'}
         \Biggl\{
(\tilde{T}^h_{\epsilon_s}-\tilde{T}^h_{\epsilon_s'})
\begin{pmatrix}
\Sigma^{s}_{n,\omega_s}\\
  \Sigma^{s}_{a,\omega_s}
\end{pmatrix}
+e^2DA_{s\omega_0}A^{pr}_{\omega}
\Biggl[
  \tilde{T}^h_{\epsilon_{-s}}
  \begin{pmatrix}
    g^+_{\epsilon_{-s}}-g^-_{\epsilon_{-s}}\\
    f^+_{\epsilon_{-s}}-f^-_{\epsilon_{-s}}\end{pmatrix}
+  \tilde{T}^h_{\epsilon'_{-s}}
  \begin{pmatrix}
    g^+_{\epsilon_{-s}'}-g^-_{\epsilon_{-s}'}\\
    f^+_{\epsilon_{-s}'}-f^-_{\epsilon_{-s}'}\end{pmatrix}
\\&-  \tilde{T}^h_{\epsilon_s'}
  \begin{pmatrix}
    g^+_{\epsilon_{-s}}
    +g^+_{\epsilon_{-s}'}\\
    f^+_{\epsilon_{-s}}
    +f^+_{\epsilon_{-s}'}\end{pmatrix}
+  \tilde{T}^h_{\epsilon_s}
  \begin{pmatrix}
    g^-_{\epsilon_{-s}}
    +g^-_{\epsilon_{-s}'}\\
    f^-_{\epsilon_{-s}}
    +f^-_{\epsilon_{-s}'}\end{pmatrix}
  \Biggr]
+\alpha
\sum_{\pm}
\Biggl[
  \begin{pmatrix}
    g^+_{\epsilon_s\pm\omega_0}
    -g^-_{\epsilon_s'\pm\omega_0}\\
    f^+_{\epsilon_s\pm\omega_0}
    -f^-_{\epsilon_s'\pm\omega_0}\end{pmatrix}
  z^{s(a)}_{\epsilon_s\pm\omega_0,\epsilon_s'\pm\omega_0}
\\&  +
  (\tilde{T}^h_{\epsilon_s'\pm\omega_0}-\tilde{T}^h_{\epsilon_s'})
  \begin{pmatrix}
    g^+_{\epsilon_s\pm\omega_0}
    -g^+_{\epsilon_s'\pm\omega_0}\\
    f^+_{\epsilon_s\pm\omega_0}
    -f^+_{\epsilon_s'\pm\omega_0}\end{pmatrix}
    y^{s,+}_{\epsilon_s\pm\omega_0,\epsilon_s'\pm\omega_0}
-
  (\tilde{T}^h_{\epsilon_s\pm\omega_0}-\tilde{T}^h_{\epsilon_s})
  \begin{pmatrix}
    g^-_{\epsilon_s\pm\omega_0}
    -g^-_{\epsilon_s'\pm\omega_0}\\
    f^-_{\epsilon_s\pm\omega_0}
    -f^-_{\epsilon_s'\pm\omega_0}\end{pmatrix}  
  y^{s,-}_{\epsilon_s\pm\omega_0,\epsilon_s'\pm\omega_0}
  \Biggr]
\Biggr\}
  \end{split}
\label{eq:eqforzsa}
\end{equation}
with
\begin{equation}
  \hat{M}^{+,\pm}_{\epsilon_s,\epsilon_s'}:=
    \frac{1}{\tilde{\zeta}^+_{\epsilon_s} +\tilde{\zeta}^{\pm}_{\epsilon_s'}}
    \begin{pmatrix}
      1+g_{\epsilon_s}^+ g_{\epsilon_s'}^{\pm}
      +f_{\epsilon_s}^+ f_{\epsilon_s'}^{\pm}
      &
            g_{\epsilon_s}^+ f_{\epsilon_s'}^{\pm}
            +f_{\epsilon_s}^+ g_{\epsilon_s'}^{\pm}
            \\
                        g_{\epsilon_s}^+ f_{\epsilon_s'}^{\pm}
            +f_{\epsilon_s}^+ g_{\epsilon_s'}^{\pm}
            &
      -1+g_{\epsilon_s}^+ g_{\epsilon_s'}^{\pm}
      +f_{\epsilon_s}^+ f_{\epsilon_s'}^{\pm}            
    \end{pmatrix}
\end{equation}
and
$\tilde{\zeta}^{\pm}_{\epsilon}:=
\sqrt{(\tilde{\Delta}^{\pm})^2-(\tilde{\epsilon}^{\pm})^2}$.
Using the relations
$\hat{g}^{\pm}_{-\epsilon}=
-\hat{\tau}_3\hat{g}^{\mp}_{\epsilon}\hat{\tau}_3$,
$\tilde{T}^h_{-\epsilon}=-\tilde{T}^h_{\epsilon}$,
$y^{s,\pm}_{-\epsilon,-\epsilon'}=-y^{s,\mp}_{\epsilon',\epsilon}$, and
$z^{s(a)}_{-\epsilon,-\epsilon'}=-z^{s(a)}_{\epsilon',\epsilon}$, 
it can be shown that  $\Sigma_{n,\omega_s}^s=0$
and 
\begin{equation}
    \Sigma_{a,\omega_s}^{s}
  =
  -p\int_{-\infty}^{\infty}\frac{d\epsilon}{2\pi i}
    \left[
    z^{s(a)}_{\epsilon_s,\epsilon_s'}
(f^+_{\epsilon_s}-f^-_{\epsilon_s'})
    +
    \tilde{T}^h_{\epsilon_s'}
    y^{s,+}_{\epsilon_s,\epsilon_s'}
(f^+_{\epsilon_s}-f^+_{\epsilon_s'})
        -
        \tilde{T}^h_{\epsilon_s}
        y^{s,-}_{\epsilon_s,\epsilon_s'}
(f^-_{\epsilon_s}-f^-_{\epsilon_s'})
        \right].
    \label{eq:siganomalous}
\end{equation}

The terms proportional to $\alpha$
on the right-hand sides of Eqs. (\ref{eq:eqforyspm})
and (\ref{eq:eqforzsa}) can be omitted
in the case of 
$\alpha\ll |\omega-\omega_0|$.
(This is in contrast to the case of $x^{(a)}_{\epsilon}$ in which
the finite inelastic scattering is important.)
Then Eq. (\ref{eq:siganomalous}) is written as 
\begin{equation}
  \begin{split}
&  \Sigma^s_{a,\omega_s}=
    -e^2DA_{s\omega_0}A^{pr}_{\omega}
    \frac{p}{L_{\omega_s}}
  \int_{-\infty}^{\infty}\frac{d\epsilon}{2\pi i}
  \Bigl\{
        \left(\tilde{T}^h_{\epsilon_s}-\tilde{T}^h_{\epsilon_s'}\right)
N_{\epsilon_s,\epsilon_s'}^{+,-}
+\tilde{T}^h_{\epsilon_s'}N^{+,+}_{\epsilon_s,\epsilon_s'}
- \tilde{T}^h_{\epsilon_s} N^{-,-}_{\epsilon_s,\epsilon_s'}
\\&+
\left(  \tilde{T}^h_{\epsilon_{-s}}-\tilde{T}^h_{\epsilon_s}\right)
     \left[
\left(\hat{M}^{+,-}_{\epsilon_s,\epsilon_s'}\right)_{12}
        (    g^+_{\epsilon_{-s}}-g^-_{\epsilon_{-s}})
        +
\left(\hat{M}^{+,-}_{\epsilon_s,\epsilon_s'}\right)_{22}
(    f^+_{\epsilon_{-s}}-f^-_{\epsilon_{-s}})\right]
\\&+  \left(\tilde{T}^h_{\epsilon_{-s}'}-\tilde{T}^h_{\epsilon_s'}\right)
\left[\left(\hat{M}^{+,-}_{\epsilon_s,\epsilon_s'}\right)_{12}
(    g^+_{\epsilon_{-s}'}-g^-_{\epsilon_{-s}'})
+\left(\hat{M}^{+,-}_{\epsilon_s,\epsilon_s'}\right)_{22}
(    f^+_{\epsilon_{-s}'}-f^-_{\epsilon_{-s}'})\right]
\Bigr\}
  \end{split}
  \label{eq:ampsigma}
\end{equation}
with
\begin{equation}
L_{\omega_s}:=
  1+p\int_{-\infty}^{\infty}\frac{d\epsilon}{2\pi i}
  \left[
    \left(\tilde{T}^h_{\epsilon_s}-\tilde{T}^h_{\epsilon_s'}\right)
\left(\hat{M}_{\epsilon_s,\epsilon_s'}^{+,-}\right)_{22}
+   \tilde{T}^h_{\epsilon_s'}
\left(\hat{M}_{\epsilon_s,\epsilon_s'}^{+,+}\right)_{22}
-   \tilde{T}^h_{\epsilon_s}
\left(\hat{M}_{\epsilon_s,\epsilon_s'}^{-,-}\right)_{22}
\right],
  \label{eq:ampdenom}
  \end{equation}
\begin{equation}
  N^{\pm,\mp}_{\epsilon_s,\epsilon_s'}:=
    \left(\hat{M}^{\pm,\mp}_{\epsilon_s,\epsilon_s'}\right)_{12}
        (g^{\pm}_{\epsilon_{-s}}+g^{\mp}_{\epsilon_{-s}'})
+\left(\hat{M}^{\pm,\mp}_{\epsilon_s,\epsilon_s'}\right)_{22}
(    f^{\pm}_{\epsilon_{-s}}    +f^{\mp}_{\epsilon_{-s}'})
\end{equation}
and $\left(\hat{M}\right)_{ij}$ means the $(i,j)$ component
of matrix $\hat{M}$.
The denominator of $\Sigma_{a,\omega_s}^s$ ($L_{\omega_s}$)  represents
the amplitude mode.

\section{Nonequilibrium Steady State}

\subsection{Conductivity}

The current is written, with the use of Eqs. (\ref{eq:approxgpmd})
and (\ref{eq:approxgad}), as 
\begin{equation}
  j_{\omega}=\frac{emk_F}{2\pi}\int\frac{d\epsilon}{4\pi i}
  \int_{FS}v_k^2{\rm Tr}
  \left[T^h_{\epsilon-\omega}\hat{g}^{+'}_{\epsilon,\epsilon-\omega}
    -T^h_{\epsilon}\hat{g}^{-'}_{\epsilon,\epsilon-\omega}
    +\hat{g}^{(a)'}_{\epsilon,\epsilon-\omega}\right].
\end{equation}
The conductivity with the linear response to
the probe field is obtained from this current as 
\begin{equation}
  \sigma_{\omega}=\sigma^{(0)}_{\omega}+\sigma^{vc}_{\omega}.
  \label{eq:condsum}
\end{equation}
Here, by using Eqs. (\ref{eq:gpmquasi}), (\ref{eq:gaquasi}),
(\ref{eq:definitionxa}), (\ref{eq:definitionyp}),
and (\ref{eq:definitionzp}),
\begin{equation}
    \frac{\sigma^{(0)}_{\omega}}{\sigma_0}=
    \frac{1}{4\omega}\int d\epsilon
       \frac{1}{2}{\rm Tr}
\left[
\tilde{T}^h_{\epsilon-\omega/2}\hat{g}^+_{\epsilon+\omega/2}
           (\hat{g}^+_{\epsilon-\omega/2}-\hat{g}^-_{\epsilon-\omega/2})
           +\tilde{T}^h_{\epsilon+\omega/2}
           (\hat{g}^+_{\epsilon+\omega/2}-\hat{g}^-_{\epsilon+\omega/2})
           \hat{g}^-_{\epsilon-\omega/2}\right]
\label{eq:cond0}
\end{equation}
and
\begin{equation}
  \begin{split}
    &    \frac{\sigma^{vc}_{\omega}}{\sigma_0}=
\sum_{s=\pm 1}\frac{A_{-s\omega_0}}{A^{pr}_{\omega}}
    \frac{1}{4\omega}\int d\epsilon
       \frac{1}{2}{\rm Tr}
\Bigl\{
          \left[\tilde{T}^h_{\epsilon_s'}
               (\hat{g}^+_{\epsilon_{-s}}+\hat{g}^-_{\epsilon'_{-s}})
+\tilde{T}^h_{\epsilon_{-s}'}
  (\hat{g}^+_{\epsilon'_{-s}}-\hat{g}^-_{\epsilon'_{-s}})
\right]           
(  \hat{g}^+_{\epsilon_s}-\hat{g}^+_{\epsilon_s'})
y^{s,+}_{\epsilon_s,\epsilon_s'}
\\&
+
\left[
  \tilde{T}^h_{\epsilon_{-s}}
 (    \hat{g}^+_{\epsilon_{-s}}-\hat{g}^-_{\epsilon_{-s}})
  -\tilde{T}^h_{\epsilon_s}
(    \hat{g}^+_{\epsilon_{-s}}+\hat{g}^-_{\epsilon'_{-s}})
\right]
(  \hat{g}^-_{\epsilon_s}-\hat{g}^-_{\epsilon'_s})
y^{s,-}_{\epsilon_s,\epsilon'_s}
+
(  \hat{g}^+_{\epsilon_{-s}}+\hat{g}^-_{\epsilon'_{-s}})
(  \hat{g}^+_{\epsilon_s}-\hat{g}^-_{\epsilon'_s})
z^{s(a)}_{\epsilon_s,\epsilon'_s}
         \Bigr\}
  \end{split}
  \label{eq:condvc}
\end{equation}
where $\sigma_0=e^2mk_FD/\pi^2$ is the conductivity
in the normal state.
[$\epsilon_s$ and $\epsilon_s'$ are
the same as those in Eq. (\ref{eq:defepssd}).]
If we put $x^{(a)}_{\epsilon}=0$
(replace $\tilde{T}^h_{\epsilon}$ by $T^h_{\epsilon}$)
in Eq. (\ref{eq:cond0}),
$\sigma_{\omega}^{(0)}$ reduces to
the Mattis--Bardeen formula for conductivity.~\cite{mattis}
There is only the $\sigma^{(0)}_{\omega}$ term
in ac conductivity for the nonequilibrium state
under a static voltage.~\cite{catelani}
The vertex correction term
$\sigma^{vc}_{\omega}$
exists under microwave irradiation and gives
a large value to the absorption spectrum,
but this term was not taken into account in
previous studies.~\cite{semenov}

\subsection{Variation of distribution function}

Equation (\ref{eq:eqforxa}) determines
$x^{(a)}_{\epsilon}$, which is the nonequilibrium correction
to the distribution function, as shown in
the definition of $\tilde{T}^h_{\epsilon}$
[Eq. (\ref{eq:deftilth})].
From Eqs. (\ref{eq:deftileps}) and (\ref{eq:deftildel}),
Eq. (\ref{eq:eqforxa}) is rewritten as 
\begin{equation}
\hat{W}
  \begin{pmatrix}
    \vdots
    \\
        x_{1}
    \\
    x_0
    \\
    x_{-1}
    \\
    \vdots
  \end{pmatrix}
  +
    \begin{pmatrix}
    \vdots
    \\
    d_{1}
    \\
 d_0
    \\
d_{-1}
    \\
    \vdots
  \end{pmatrix}
  =
  -\hat{W}
  \begin{pmatrix}
    \vdots
    \\
t_{1}
    \\
t_{0}
    \\
t_{-1}
    \\
    \vdots
  \end{pmatrix}.
        \label{eq:tridiageq}
\end{equation}
Here, $x_j:=x^{(a)}_{\epsilon+j\omega_0}$,
$t_j:=T^h_{\epsilon+j\omega_0}$,
and
\begin{equation}
  \hat{W}:=
      \begin{pmatrix}
    \ddots&\ddots&0&\cdots&\cdots
    \\
    -a_{1}
    &a_{1}+a_{0}
    &-a_{0}&0&\cdots
    \\
    0&-a_{0}&
    a_{0}+a_{-1}
    &
    -a_{-1}&0
    \\
    \cdots&0&
    -a_{-1}&
    a_{-1}+a_{-2}
    &
    -a_{-2}
    \\
    \cdots&\cdots&0&\ddots&\ddots
      \end{pmatrix}
      \label{eq:tridiagw}
\end{equation}
with
\begin{equation}
a_j:=
4\alpha
\left(
{\rm Im}g^+_{\epsilon+j\omega_0}
  {\rm Im}g^+_{\epsilon+(j+1)\omega_0}
  +
  {\rm Im}f^+_{\epsilon+j\omega_0}
  {\rm Im}f^+_{\epsilon+(j+1)\omega_0}\right),
  \label{eq:defaj}
\end{equation}
\begin{equation}
  d_{j}=d_j^{\gamma}x_j+d_j^{(a)},
    \label{eq:defdj}
\end{equation}
\begin{equation}
    d_{j}^{\gamma}:=4\left(
           {\rm Im}\Sigma^+_{n,\epsilon+j\omega_0}
  {\rm Im}g^+_{\epsilon+j\omega_0}
+  
     {\rm Im}\Sigma^+_{a,\epsilon+j\omega_0}
{\rm Im}f^+_{\epsilon+j\omega_0}
\right),
\label{eq:defdjgam}
\end{equation}
and
\begin{equation}
    d^{(a)}_{j}:=2i\left(
        \Sigma^{(a)}_{n,\epsilon+j\omega_0}
  {\rm Im}g^+_{\epsilon+j\omega_0}
+  
\Sigma^{(a)}_{a,\epsilon+j\omega_0}
{\rm Im}f^+_{\epsilon+j\omega_0}
\right).
\label{eq:defdja}
\end{equation}
In the absence of the inelastic scattering effect ($d_j=0$),
the solution to Eq. (\ref{eq:tridiageq})
is given by $t_j+x_j=t_{j\pm 1}+x_{j\pm 1}$, and this leads to
$\tilde{T}^h_{\epsilon}=T^h_{\epsilon}+
x^{(a)}_{\epsilon}=0$~\cite{memo1}
because $\tilde{T}^h_{-\epsilon}=
-\tilde{T}^h_{\epsilon}$.
This means that 
the effective temperature [$T_{eff}$
  if we write $\tilde{T}^h_{\epsilon}={\rm tanh}(\epsilon/2T_{eff})$]
becomes infinity ($T_{eff}\to \infty$), which
is the result under monochromatic microwave irradiation
with no energy dissipation.
(The divergence of perturbative expansions
by external fields
in the absence of inelastic scattering~\cite{yudson,jujo17}
is related to this.)
On the other hand, if we assume
the inelastic scattering effect to be predominant over
the irradiation effect ($|{\rm Im}\Sigma|\gg
\alpha|{\rm Im}g|$),
the solution to Eq. (\ref{eq:tridiageq}) is
approximately given by $x_j\simeq 0$.
This gives $\tilde{T}^h_{\epsilon}\simeq
T^h_{\epsilon}={\rm tanh}(\epsilon/2T)$,
and then the effective temperature takes the same value
as that in the equilibrium state ($T_{eff}\simeq T$),
which indicates that the energy injected by the irradiation
is quickly dissipated by inelastic scattering.

\subsection{Inelastic scattering}

We take the interaction between electrons
and acoustic phonons as the inelastic scattering effect.
The reason for considering acoustic phonons as a factor of the inelastic
scattering is that low-energy phonons are important
for the energy dissipation in order to maintain
a nonequilibrium steady state.
The self-energy terms due to this inelastic scattering 
in Eqs. (\ref{eq:sigeppm}) and (\ref{eq:sigepa})
are written as~\cite{kopnin}
\begin{equation}
  \hat{\Sigma}^{\gamma,\pm}_{\epsilon}
  =\frac{p'}{(2v_sk_F)^2}
  \int d\epsilon'(\epsilon-\epsilon')|\epsilon-\epsilon'|
  \hat{\tau}_3
  \left[{\rm coth}\frac{\epsilon-\epsilon'}{2T}\hat{g}^{\pm}_{k'}
    \pm\frac{1}{2}
    \left({\rm tanh}\frac{\epsilon'}{2T}+x^{(a)}_{\epsilon'}
    \right)
    \left(\hat{g}^+_{k'}-\hat{g}^-_{k'}\right)
    \right]\hat{\tau}_3
  \label{eq:defsigacphpm}
\end{equation}
and
\begin{equation}
  \begin{split}
    \hat{\Sigma}^{\gamma(a)}_{\epsilon}
 &=
  \frac{p'}{(2v_sk_F)^2}
  \int d\epsilon'(\epsilon-\epsilon')|\epsilon-\epsilon'|
  \hat{\tau}_3(\hat{g}^+_{\epsilon'}-\hat{g}^-_{\epsilon'})
  \hat{\tau}_3
  \left({\rm coth}\frac{\epsilon-\epsilon'}{2T}
  -{\rm tanh}\frac{\epsilon}{2T}\right)
  x^{(a)}_{\epsilon'}
  \end{split}
    \label{eq:defsigacpha}
\end{equation}
with $p'$ being the effective coupling constant between
electrons and acoustic phonons
[$p'$ is different from $p$
  because $p$ in the gap equation Eq. (\ref{eq:tau1gapeq})
is considered to originate from the optical phonons]
and $v_s$ being
the velocity of sound (the value of this quantity is fixed to
$v_sk_F=8.0$ in the numerical calculation below).
Here, it is assumed that phonons are kept in equilibrium
as a heat bath. Since the distribution function of phonons is
integrated by energy,
this approximation is considered
to be valid when we treat the steady-state phenomena
rather than transient phenomena.
Using Eqs. (\ref{eq:defsigacphpm}) and (\ref{eq:defsigacpha}),
Eq. (\ref{eq:defdj}) with $j=0$ is written as
\begin{equation}
  \begin{split}
&  d_0
  =
    \frac{p'}{(2v_sk_F)^2}
  \int d\epsilon'(\epsilon-\epsilon')|\epsilon-\epsilon'|
[  (g^+_{\epsilon}-g^-_{\epsilon})
  (g^+_{\epsilon'}-g^-_{\epsilon'})
  -
    (f^+_{\epsilon}-f^-_{\epsilon})
  (f^+_{\epsilon'}-f^-_{\epsilon'})]
\\&
\times
\left[
  \left({\rm coth}\frac{\epsilon-\epsilon'}{2T}
  -{\rm tanh}\frac{\epsilon}{2T}\right)
  x^{(a)}_{\epsilon'}
  -
  \left({\rm coth}\frac{\epsilon-\epsilon'}{2T}
  +{\rm tanh}\frac{\epsilon'}{2T}
  +x^{(a)}_{\epsilon'}
  \right)
  x^{(a)}_{\epsilon}
    \right].
  \end{split}
\end{equation}

\section{Results of Numerical Calculations}

In this section we show the results of
the numerical calculation for the absorption spectrum
${\rm Re}\sigma_{\omega}$.
The superconducting gap 
in the absence of 
microwave irradiation ($\alpha=0$) at $T=0$
is taken to be the unit of energy 
  and denoted as $\Delta_0$.
The unit of temperature is given by
the transition temperature
$T_c$, which is also obtained by the gap equation
in the case of $\alpha=0$.
Although the values of $\Delta$ and $T_c$ for
$\alpha\ne 0$ are different from those for $\alpha= 0$
especially around the superconducting transition,
numerical calculations in this work are performed
in the range of $\alpha$ and $T$ where
the dependence of $\Delta$ on $\alpha$ is small
[the variation of the gap equation Eq. (\ref{eq:tau1gapeq})
is smaller than 0.01].
The obtained result indicates that
the absorption spectrum shows a large suppression
even in the case of such small values of $\alpha$.
The range of energy is taken to be $|\epsilon|\le 8\Delta_0$.

\begin{figure}
  \includegraphics[width=11.5cm]{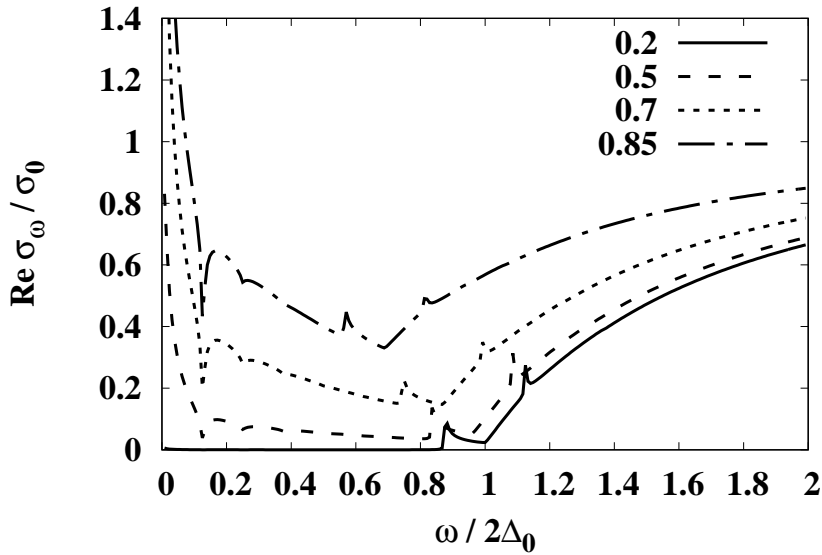}
  \caption{
    \label{fig:1}
    Dependences of ${\rm Re}\sigma_{\omega}$ on $\omega$
    for $T/T_c=0.2$, $0.5$, $0.7$, and $0.85$.
    $\alpha/\Delta_0=0.0005$,
    $\omega_0/\Delta_0=0.25$, and 
    $p'/\Delta_0=0.05$.
  }
\end{figure}

The calculated result of the absorption spectrum
[the real part of Eq. (\ref{eq:condsum})]
is shown in Fig.~\ref{fig:1}.
The absorption results from two types of
excitation: 
the thermal excitation of quasiparticles for
($\omega<2\Delta$) and 
the direct excitation for ($\omega>2\Delta$).
This is similar to the results of the Mattis--Bardeen formula~\cite{mattis}
in the linear response,
but there are two main differences in the nonequilibrium state.
One of these is the suppression of the absorption for $\omega<2\Delta$
with dip structures around $\omega\simeq n\omega_0$ ($n=1,2,...$).
This feature is caused by the
variation of the distribution function
under the external field
by $x^{(a)}_{\epsilon}$, as shown below.
Another feature is that
there are two peaks around $\omega\simeq 2\Delta\pm\omega_0$,
and these originate from the amplitude mode.

\begin{figure}
  \includegraphics[width=8.5cm]{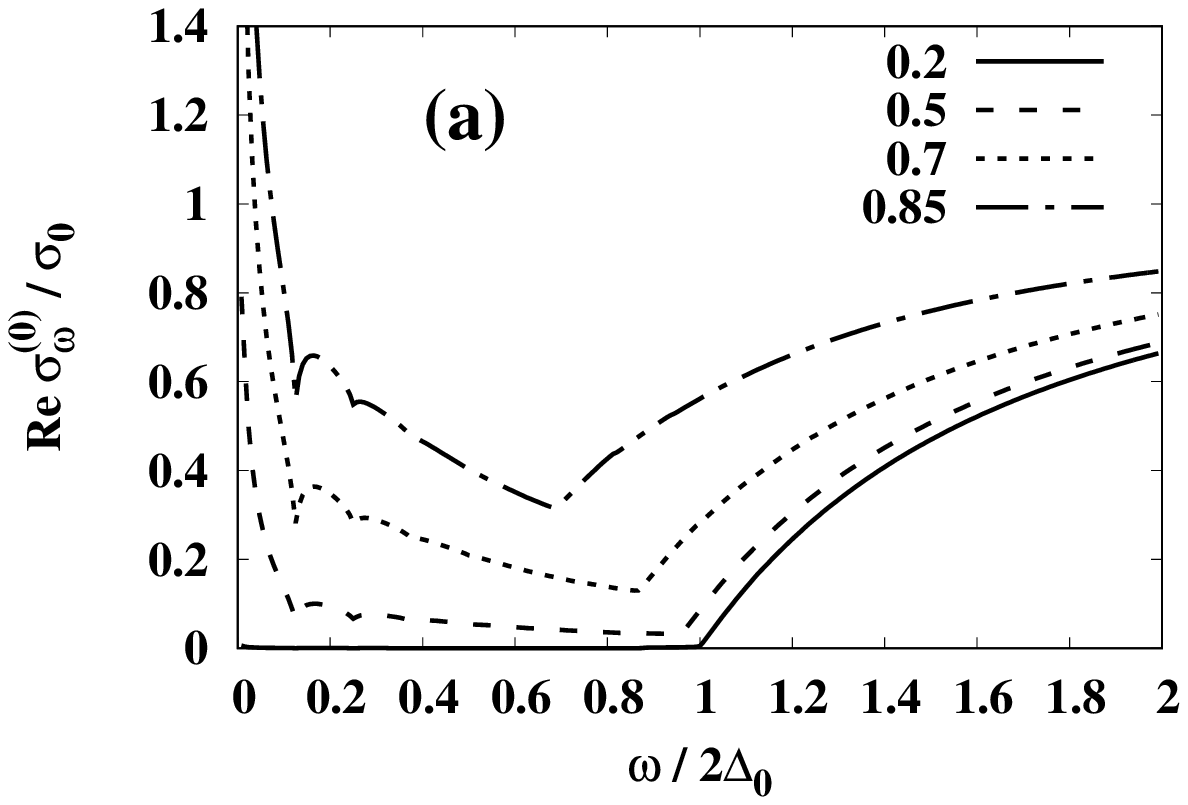}
  \includegraphics[width=8.5cm]{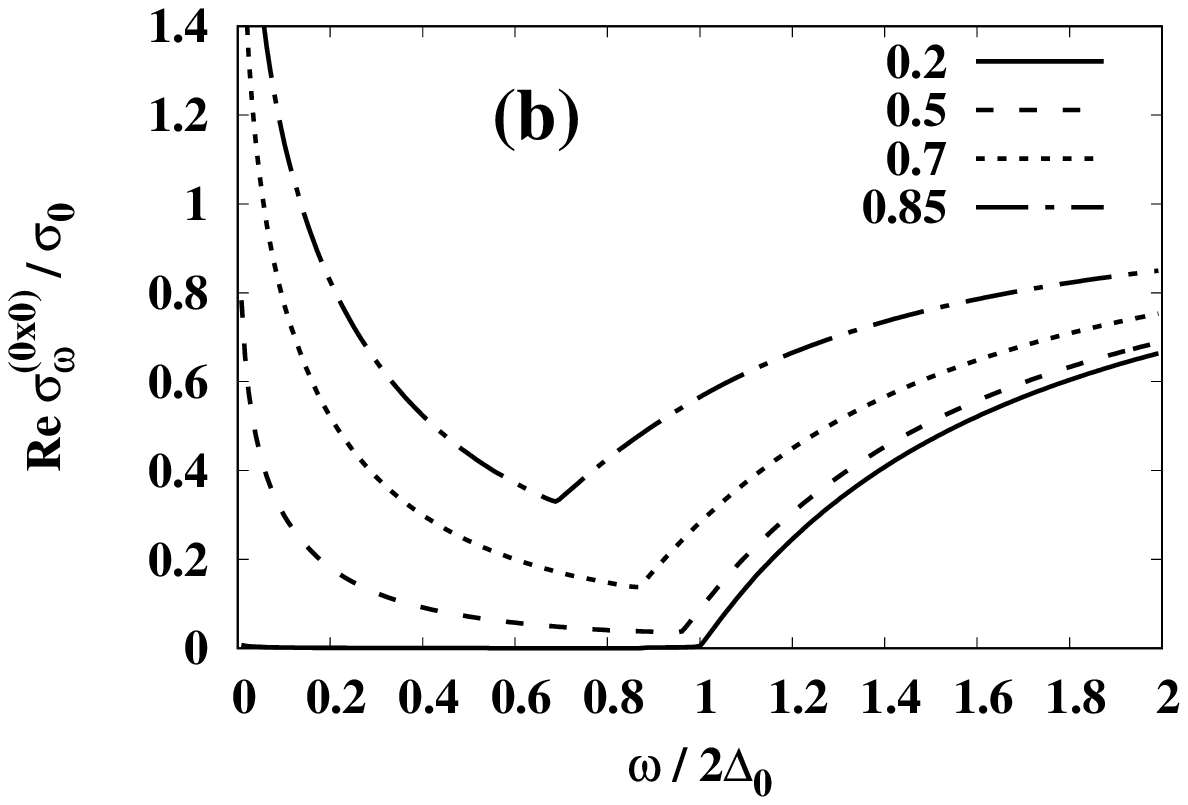}
  \includegraphics[width=8.5cm]{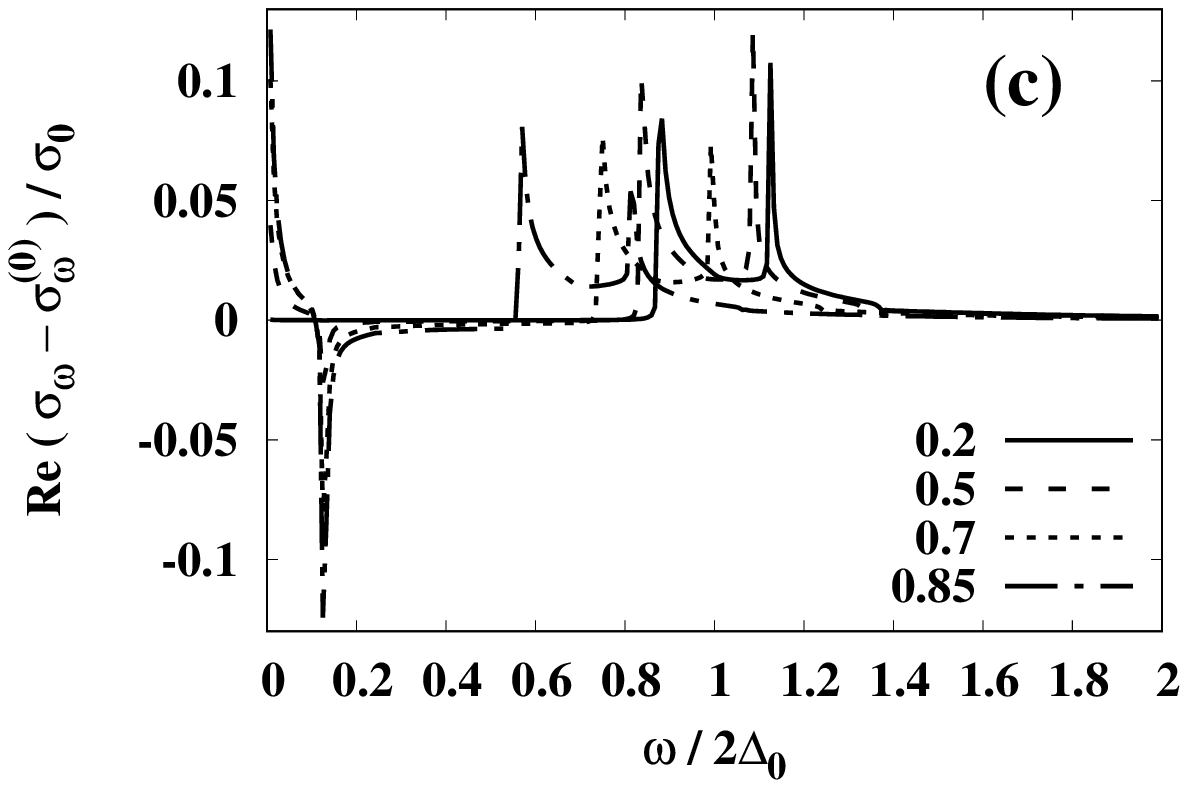}
  \includegraphics[width=8.5cm]{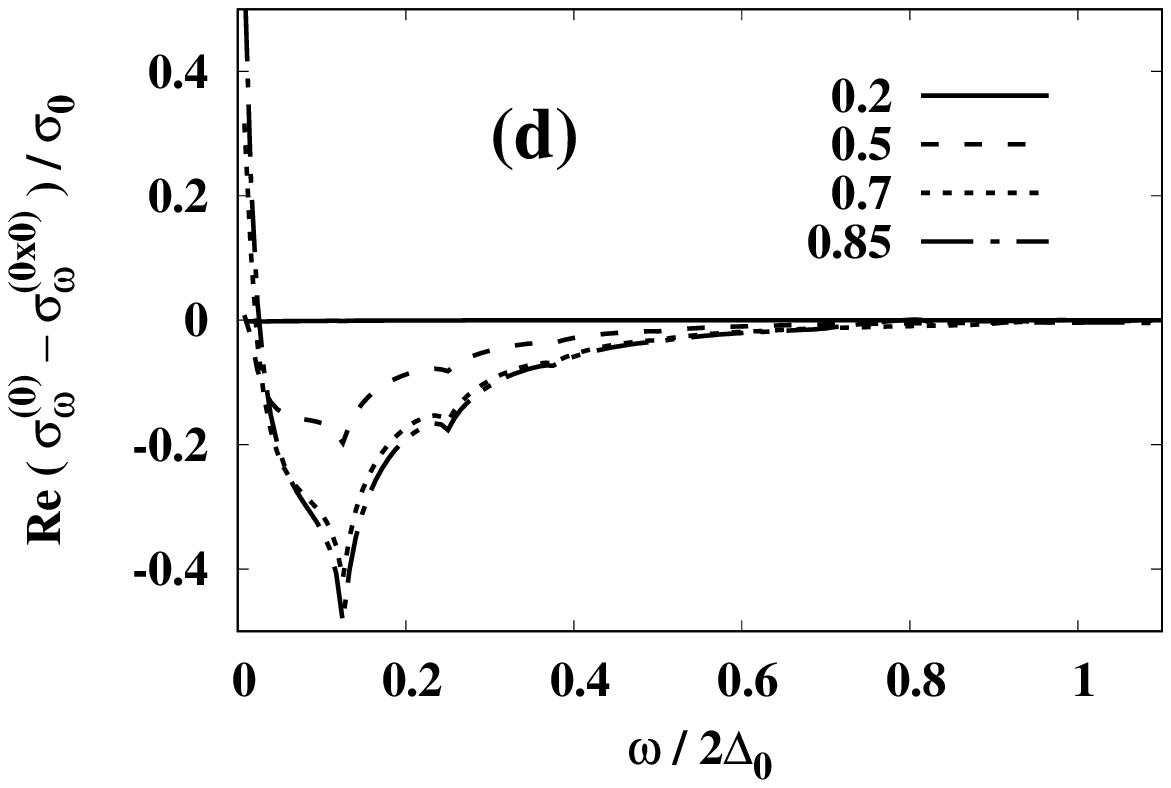}
  \caption{
    \label{fig:2}
    Dependences of (a) ${\rm Re}\sigma^{(0)}_{\omega}$,
    (b) ${\rm Re}\sigma^{(0x0)}_{\omega}$,
    (c) ${\rm Re}\left(\sigma_{\omega}-\sigma^{(0)}_{\omega}\right)$,
and
    (d) ${\rm Re}\left(\sigma_{\omega}^{(0)}-\sigma^{(0x0)}_{\omega}\right)$  
    on $\omega$
    for $T/T_c=0.2$, $0.5$, $0.7$, and $0.85$.
    $\alpha/\Delta_0=0.0005$,
    $\omega_0/\Delta_0=0.25$, and 
    $p'/\Delta_0=0.05$.
  }
\end{figure}

The absorption spectrum without vertex corrections
[the real part of Eq. (\ref{eq:cond0})]
is shown in Fig.~\ref{fig:2}(a).
In Fig. 2(a), there is no peak around $\omega\simeq 2\Delta\pm\omega_0$,
and the absorption by thermally excited quasiparticles
is suppressed, as shown in Fig. 1.
This is clearly seen by comparing this result with
the absorption in the equilibrium state
[${\rm Re}\sigma_{\omega}^{(0x0)}$], which
is shown in Fig. 2(b).
Here, $\sigma_{\omega}^{(0x0)}$ is defined
as the quantity obtained 
by replacing $\tilde{T}^h_{\epsilon}=T^h_{\epsilon}+x^{(a)}_{\epsilon}$ 
by $T^h_{\epsilon}$ in Eq. (\ref{eq:cond0}). 
  The variation of the absorption spectrum
  owing to the change of the density of states for $\alpha\ne 0$
  is small.
  When the vertex correction and the nonequilibrium correction
  to the distribution function are ignored,
  the absorption spectrum almost agrees with that for $\alpha=0$.
  This is because the nonequilibrium correction to
  the distribution function is absent in the linear response
  without the pump field.

The difference between ${\rm Re}\sigma_{\omega}$
and ${\rm Re}\sigma^{(0)}_{\omega}$
shown in Fig. 2(c) indicates
that the amplitude mode originates from
the vertex correction term [Eq. (\ref{eq:condvc})].
The positive peak of ${\rm Re}\left(\sigma_{\omega}
-\sigma^{(0)}_{\omega}\right)$ at small $\omega$ indicate the
thermally excited quasiparticle contribution
from the vertex correction term to the absorption. 
The negative peak at $\omega\simeq \omega_0$
originates from the overlap of the density of states
of $\hat{g}^{\pm}_{\epsilon_{-1}}$ and $\hat{g}^{\pm}_{\epsilon'_{-1}}$
when $\omega_{-1}=0$ in the $s=-1$ term of Eq. (\ref{eq:condvc}).
At this frequency, the probe wave overlaps with the pump wave,
and this peak results from the perturbation by the probe field,
which should originally be treated as a single wave with the pump
wave.
Figure 2(d) shows the suppression of the absorption
spectrum induced by $x^{(a)}_{\epsilon}$
which means the redistribution of quasiparticles
in the nonequilibrium state.  
(The increase around small $\omega$ is discussed below.)

\begin{figure}
  \includegraphics[width=11.5cm]{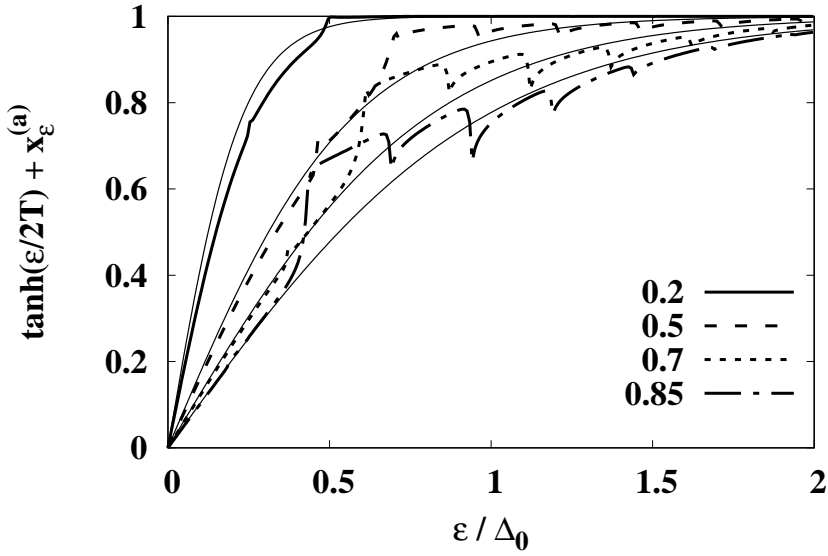}
  \caption{
    \label{fig:3}
    Dependences of
    $\tilde{T}^h_{\epsilon}={\rm tanh}(\epsilon/2T)+x^{(a)}_{\epsilon}$
 on $\epsilon$
    for $T/T_c=0.2$, $0.5$, $0.7$, and $0.85$.
    $\alpha/\Delta_0=0.0005$,
    $\omega_0/\Delta_0=0.25$, and 
    $p'/\Delta_0=0.05$.
    Thin lines are graphs of $T^h_{\epsilon}={\rm tanh}(\epsilon/2T)$
for $T/T_c=0.2$, $0.5$, $0.7$, and $0.85$ from left to right, respectively.
  }
\end{figure}

The calculated result of $x^{(a)}$ is shown in 
Fig.~\ref{fig:3}.
A steplike behavior 
with the interval $\omega_0$ exists, which is the result
of the coupling term between $x^{(a)}_{\epsilon}$
and $x^{(a)}_{\epsilon\pm\omega_0}$ in Eq. (\ref{eq:eqforxa}).
As the Fermi distribution function in the equilibrium
state is written as $1/(e^{\epsilon/T}+1)=(1-T^h_{\epsilon})/2$,
the variation of the distribution function
in the nonequilibrium state
is given by $-x^{(a)}_{\epsilon}/2$
because of the replacement of $T^h_{\epsilon}$
by $\tilde{T}^h_{\epsilon}$. 
Figure 3 shows that
$\tilde{T}^h_{\epsilon}<T^h_{\epsilon}$
for $\epsilon>\Delta+\omega_0$
and $\epsilon< \Delta-2\omega_0$,
and
there exists a region such that 
$\tilde{T}^h_{\epsilon}>T^h_{\epsilon}$
for $\Delta-2\omega_0<\epsilon<\Delta+\omega_0$.
[The positions of $\epsilon=\Delta\pm n\omega_0$ ($n=0,1,2,...$)
correspond to energies at which the graph shows dips or bends.]  
This indicates that quasiparticles in the nonequilibrium state
redistribute
[$(1-\tilde{T}^h_{\epsilon})/2$] 
from the low energy ($\Delta-2\omega_0<\epsilon<\Delta+\omega_0$)
to the high energy ($\epsilon>\Delta+\omega_0$).
(The thermalization due to the inelastic scattering
makes $\tilde{T}^h_{\epsilon}$ smaller than $T^h_{\epsilon}$
at low energies $0<\epsilon< \Delta-2\omega_0$.)
In previous studies, the distribution function
in the range of $\epsilon>\Delta$ was investigated.
Here, we show that
$\tilde{T}^h_{\epsilon}>T^h_{\epsilon}$
holds at lower energies 
when the range of $\epsilon<\Delta$ is taken into account.

\begin{figure}
  \includegraphics[width=8.5cm]{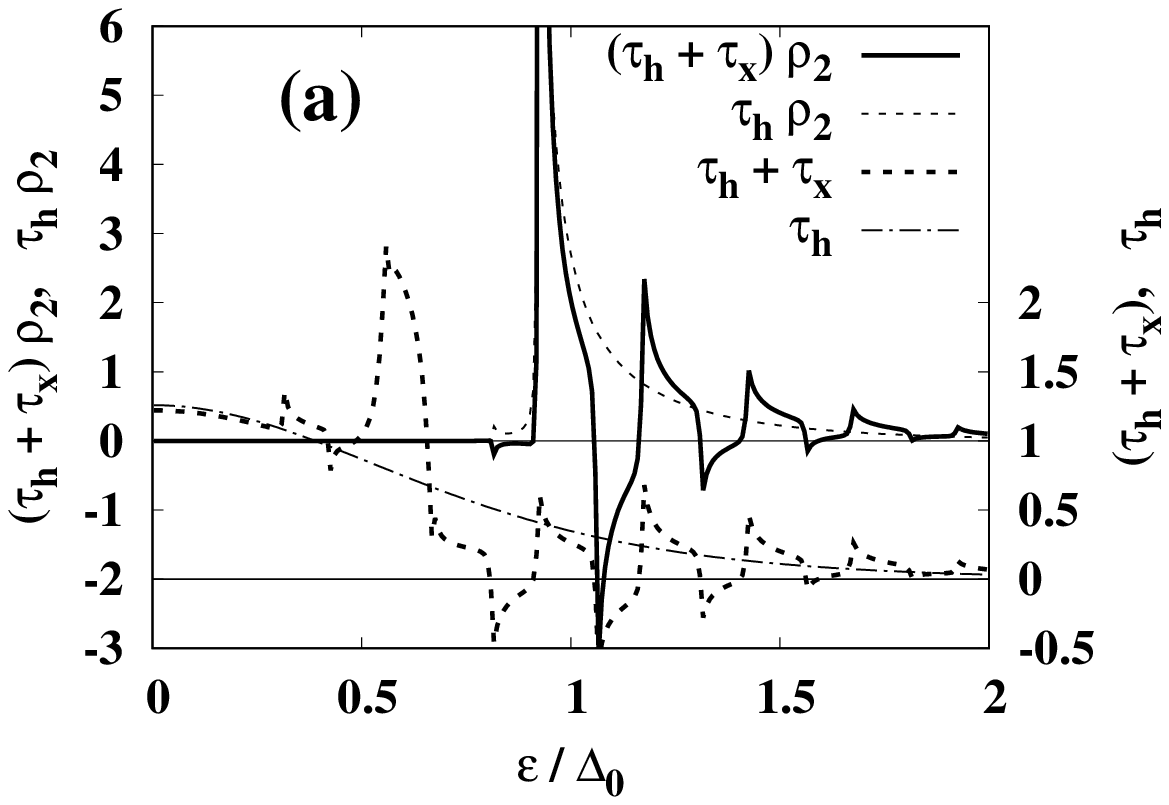}
  \includegraphics[width=8.5cm]{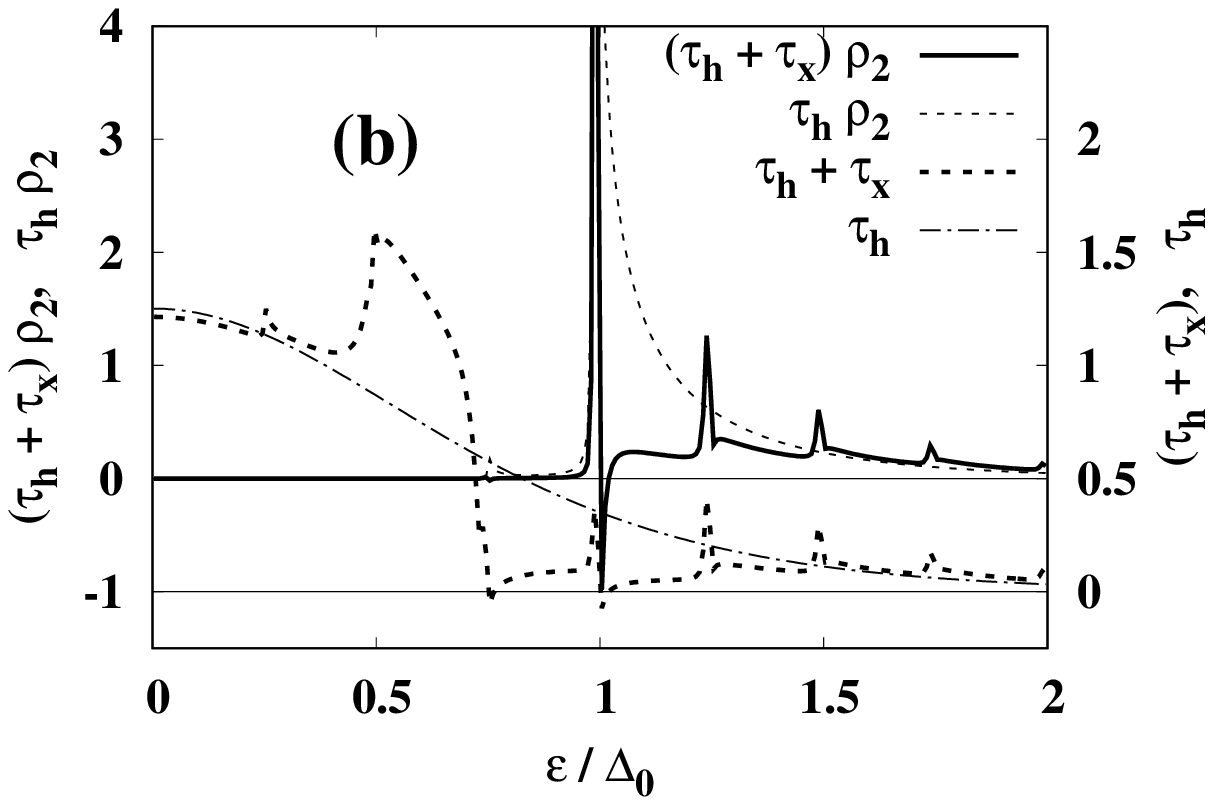}
  \includegraphics[width=8.5cm]{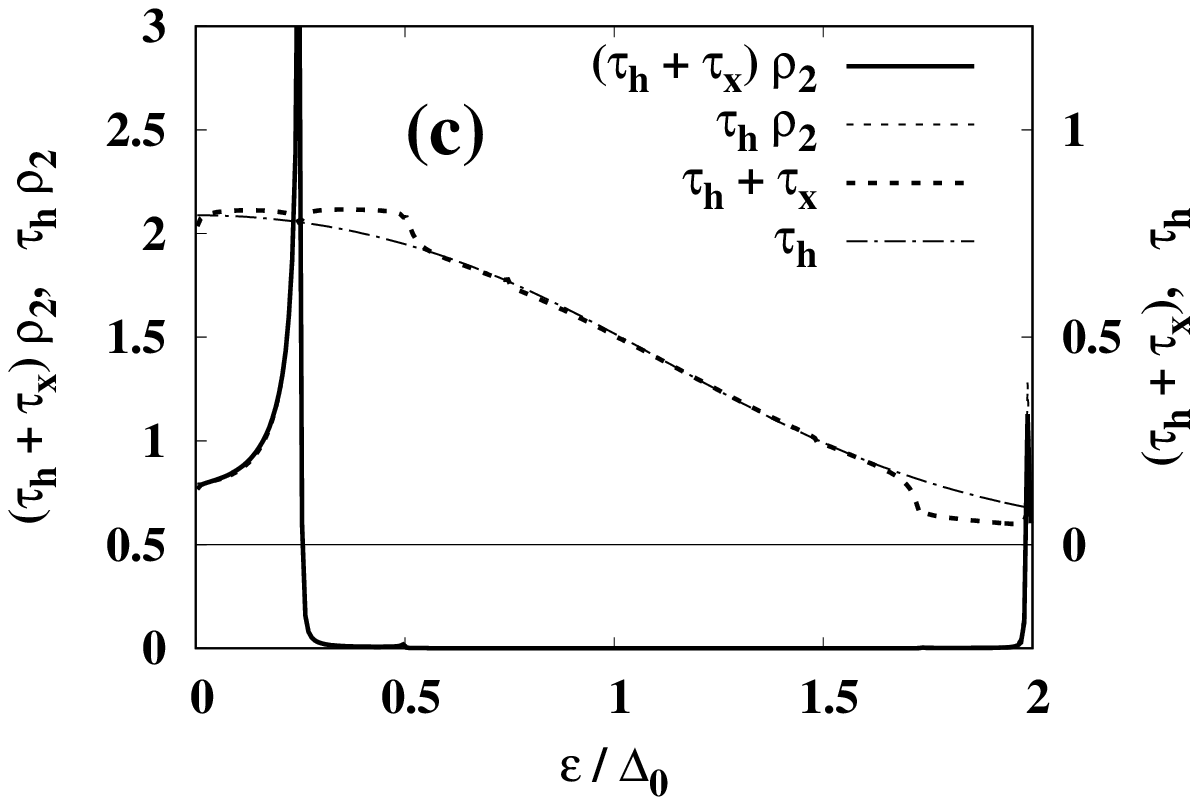}
  \includegraphics[width=8.5cm]{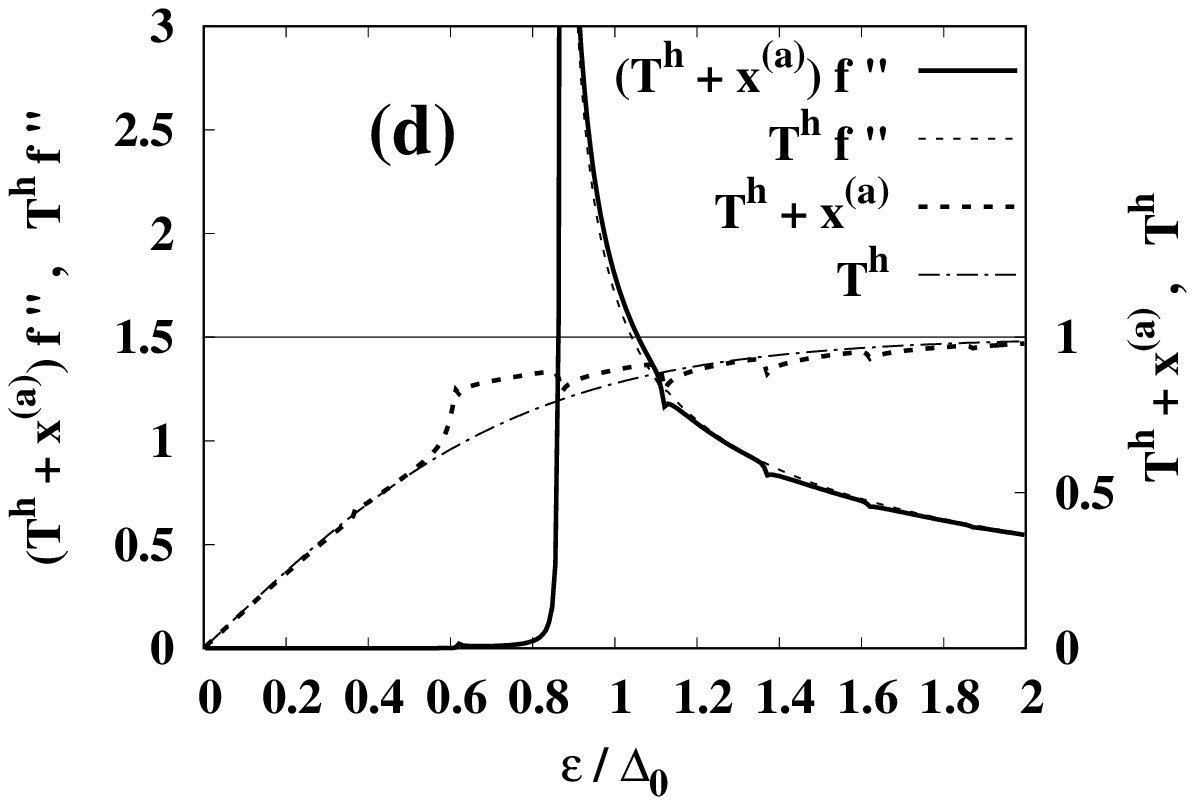}
  \caption{
    \label{fig:4}
    Dependences of 
    $\left[\tau_h(\epsilon,\omega)+\tau_x(\epsilon,\omega)\right]
    \rho_2(\epsilon,\omega)$,
    $\tau_h(\epsilon,\omega)\rho_2(\epsilon,\omega)$,
    $\tau_h(\epsilon,\omega)+\tau_x(\epsilon,\omega)$,
    and $\tau_h(\epsilon,\omega)$
    on $\epsilon$
for    (a) $\omega=\omega_0/2=0.125\Delta_0$,
(b) $\omega=\omega_0=0.25\Delta_0$,
and
(c) $\omega=2.25\Delta_0$.
(d) Dependences of
$-\tilde{T}^h_{\epsilon}{\rm Im}f^+_{\epsilon}/\Delta$ and 
$-T^h_{\epsilon}{\rm Im}f^+_{\epsilon}/\Delta$
on $\epsilon$. ($f'':=-{\rm Im}f^+_{\epsilon}/\Delta$.)
    $T/T_c=0.7$,
        $\alpha/\Delta_0=0.0005$,
    $\omega_0/\Delta_0=0.25$, and 
    $p'/\Delta_0=0.05$.
  }
\end{figure}

We introduce the following quantities to
see how this $x^{(a)}_{\epsilon}$
suppresses the absorption spectrum
in the nonequilibrium state:
\begin{equation}
\rho_{2}(\epsilon,\omega)
:={\rm Im}g^+_{\epsilon+\omega/2}{\rm Im}g^+_{\epsilon-\omega/2}
+{\rm Im}f^+_{\epsilon+\omega/2}{\rm Im}f^+_{\epsilon-\omega/2},
  \label{eq:defofrho2}
\end{equation}
\begin{equation}
\tau_h(\epsilon,\omega)
:=(T^h_{\epsilon+\omega/2}-T^h_{\epsilon-\omega/2})/\omega,
  \label{eq:defoftauh}
\end{equation}
and
\begin{equation}
  \tau_x(\epsilon,\omega)
  :=(x^{(a)}_{\epsilon+\omega/2}-x^{(a)}_{\epsilon-\omega/2})/\omega.
  \label{eq:defoftaux}
\end{equation}
With the use of these quantities, 
the absorption spectra shown
in Figs. 2(a) and 2(b) are written as 
\begin{equation}
  {\rm Re}\sigma^{(0)}_{\omega}/\sigma_0=
    (1/2)\int d\epsilon
\left[\tau_h(\epsilon,\omega)+\tau_x(\epsilon,\omega)\right]
\rho_2(\epsilon,\omega)
\label{eq:resig0bytau}
\end{equation}
and
\begin{equation}
  {\rm Re}\sigma^{(0x0)}_{\omega}/\sigma_0:=
(1/2)\int d\epsilon
    \tau_h(\epsilon,\omega)
    \rho_2(\epsilon,\omega).
    \label{eq:resig0x0bytau}
\end{equation}
Figures~\ref{fig:4}(a)--4(c) show
how
$[\tau_h(\epsilon,\omega)+\tau_x(\epsilon,\omega)]\rho_2(\epsilon,\omega)$
is suppressed compared with
$\tau_h(\epsilon,\omega)\rho_2(\epsilon,\omega)$.
The quantity $\rho_2(\epsilon>0,\omega)$
takes large values for $\epsilon>\Delta+\omega/2$
in the case of $\omega<2\Delta$ [Figs. 4(a) and 4(b)]
and for $\epsilon<-\Delta+\omega/2$
in the case of $\omega>2\Delta$ [Fig. 4(c)].
In Figs. 4(a) and 4(b),
the value of $\tau_h(\epsilon,\omega)$ is small for
this range of $\epsilon>\Delta+\omega/2$
compared with the value at its peak around $\epsilon\simeq 0$:
thus
the variation of $\tau_x(\epsilon,\omega)$ 
effectively determines the behavior of
$\tau_h(\epsilon,\omega)+\tau_x(\epsilon,\omega)$.
This leads to a large suppression of the absorption
spectrum for $\omega<2\Delta$.
In particular, the negative contribution
from $x^{(a)}_{\epsilon}$ to
$\tau_x(\epsilon,\omega)$ is enhanced for $\omega=n\omega_0$
($n=1,2,...$) 
because the negative parts of 
$x^{(a)}_{\epsilon+\omega/2}$ with the interval $\omega_0$
overlap with the positive parts of
$x^{(a)}_{\epsilon-\omega/2}$ at this frequency.
This is in contrast to the case of
$\omega>2\Delta$ in Fig. 4(c).
In this case, $\tau_h(\epsilon,\omega)$
is predominant over $\tau_x(\epsilon,\omega)$
because the latter quantity is ineffective in this range of $\omega$.
Then the nonequilibrium effect causes 
only a small change in the absorption
spectrum for $\omega>2\Delta$.
The same consideration can be applied to 
the gap equation [Eq. (\ref{eq:tau1gapeq})],
which is rewritten as 
\begin{equation}
\pi/p = 
    \int d\epsilon
    \tilde{T}^h_{\epsilon}
    {\rm Im}(-f^+_{\epsilon}/\Delta).
\end{equation}
The calculated results of
$\tilde{T}^h_{\epsilon}    {\rm Im}(-f^+_{\epsilon}/\Delta)$
and $T^h_{\epsilon}    {\rm Im}(-f^+_{\epsilon}/\Delta)$
are shown in Fig. 4(d).
The variation due to the nonequilibrium effect
($\tilde{T}^h_{\epsilon}-T^h_{\epsilon}$)
is shown to be small
compared with the equilibrium distribution
function $T^h_{\epsilon}$
in the range of $\epsilon$ where the density of states is large.
Then we can neglect the variation of $\Delta$ 
in the presence of the finite value of $\alpha$,
as noted above, although
there is a large variation in the absorption
by the thermally excited quasiparticles
even for small values of $\alpha$.

\begin{figure}
  \includegraphics[width=8.5cm]{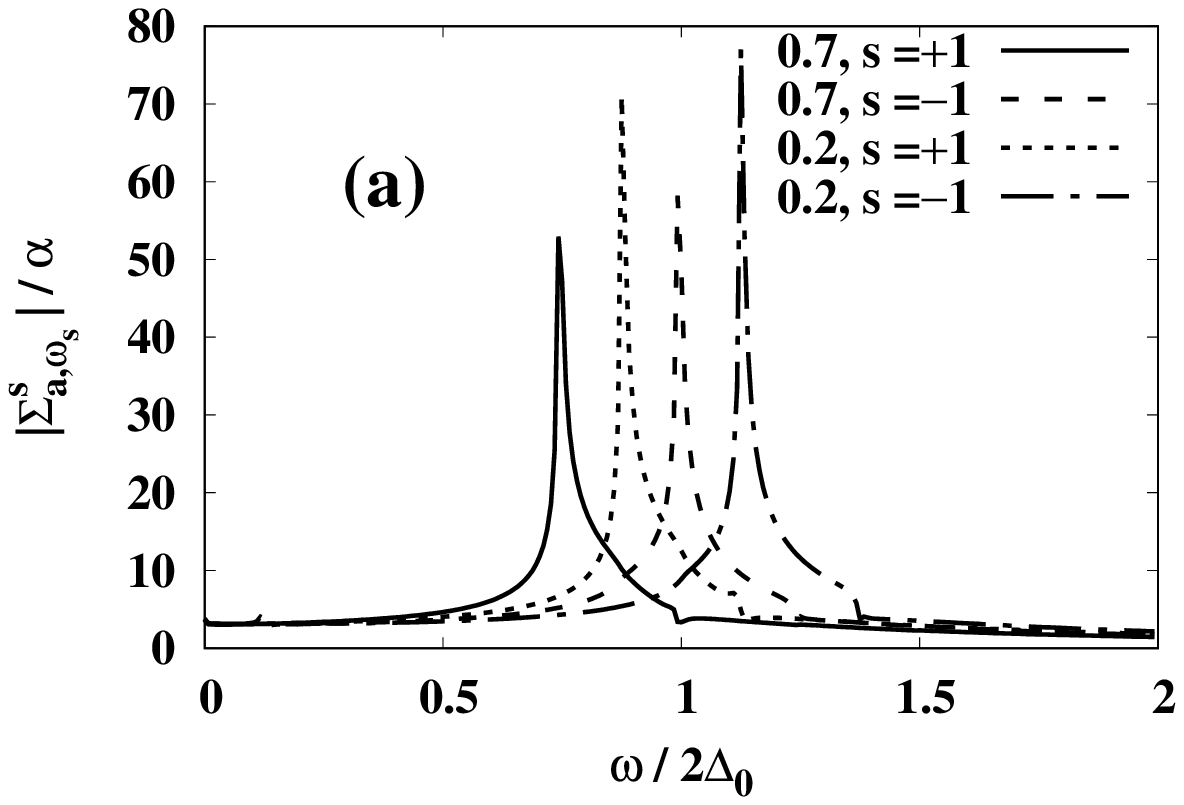}
  \includegraphics[width=8.5cm]{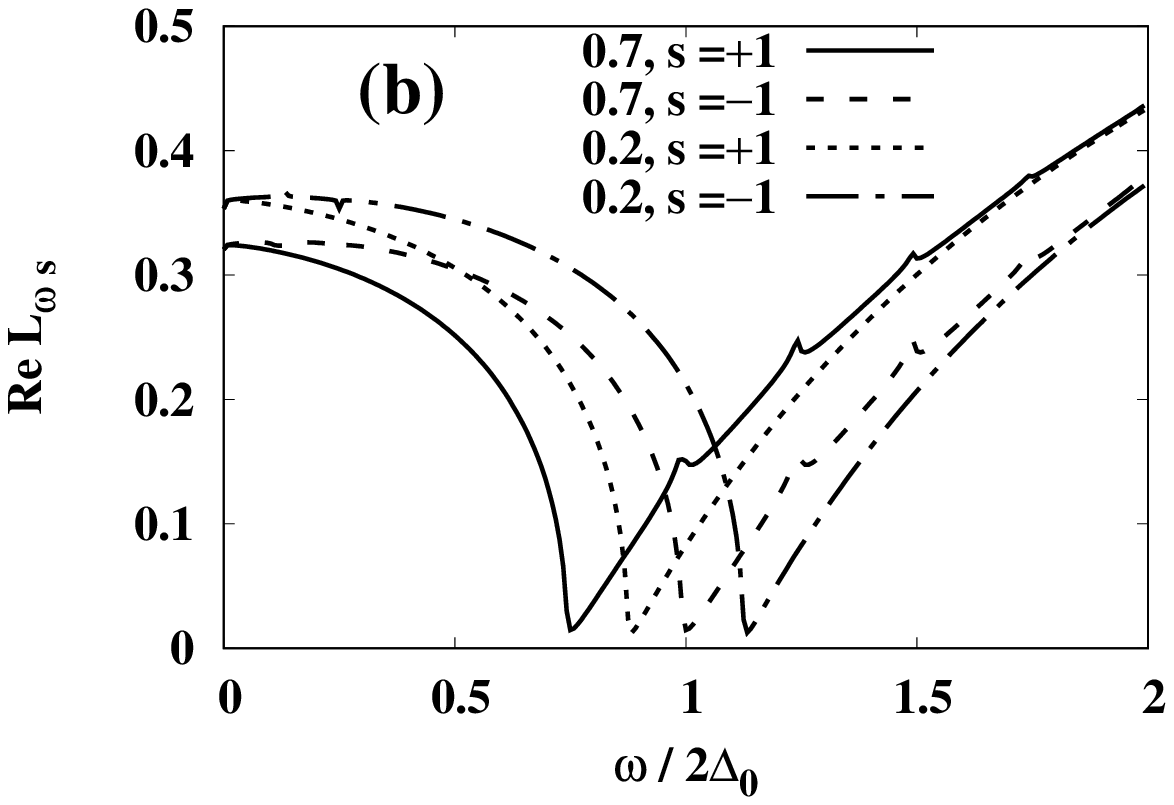}
  \caption{
    \label{fig:5}
    Dependences of (a) $|\Sigma^s_{a,\omega_s}|/\alpha$
    and (b) ${\rm Re}L_{\omega_s}$ on $\omega$
    for $T/T_c=0.2$ and $0.7$.
        $\alpha/\Delta_0=0.0005$,
    $\omega_0/\Delta_0=0.25$, and 
    $p'/\Delta_0=0.05$.
  }
\end{figure}

The vertex correction term including
the amplitude mode [Eq. (\ref{eq:ampsigma})]
and its denominator describing this mode [Eq. (\ref{eq:ampdenom})]
are shown in
Figs.~\ref{fig:5}(a) and 5(b), respectively.
Figure 5(a) shows that
at each $T/T_c$, two peaks appear:
one at $\omega< 2\Delta$
and the other at $\omega> 2\Delta$.
From Eq. (\ref{eq:defoms}),
$\omega_{\pm 1}=\omega\pm \omega_0=2\Delta$
means that 
$\omega=2\Delta\mp\omega_0$.
Then the peaks of 
$|\Sigma^{\pm 1}_{a,\omega_{\pm 1}}|$
  indicate the existence of the amplitude mode
  at these frequencies.
  This is confirmed by 
  ${\rm Re}L_{\omega}$ being minimum at the same frequencies.
The frequencies ($\omega=2\Delta\mp\omega_0$) at which 
two peaks appear 
are different between $T/T_c=0.2$ and $0.7$ because of
the dependence of $\Delta$ on $T$.

\begin{figure}
  \includegraphics[width=8.5cm]{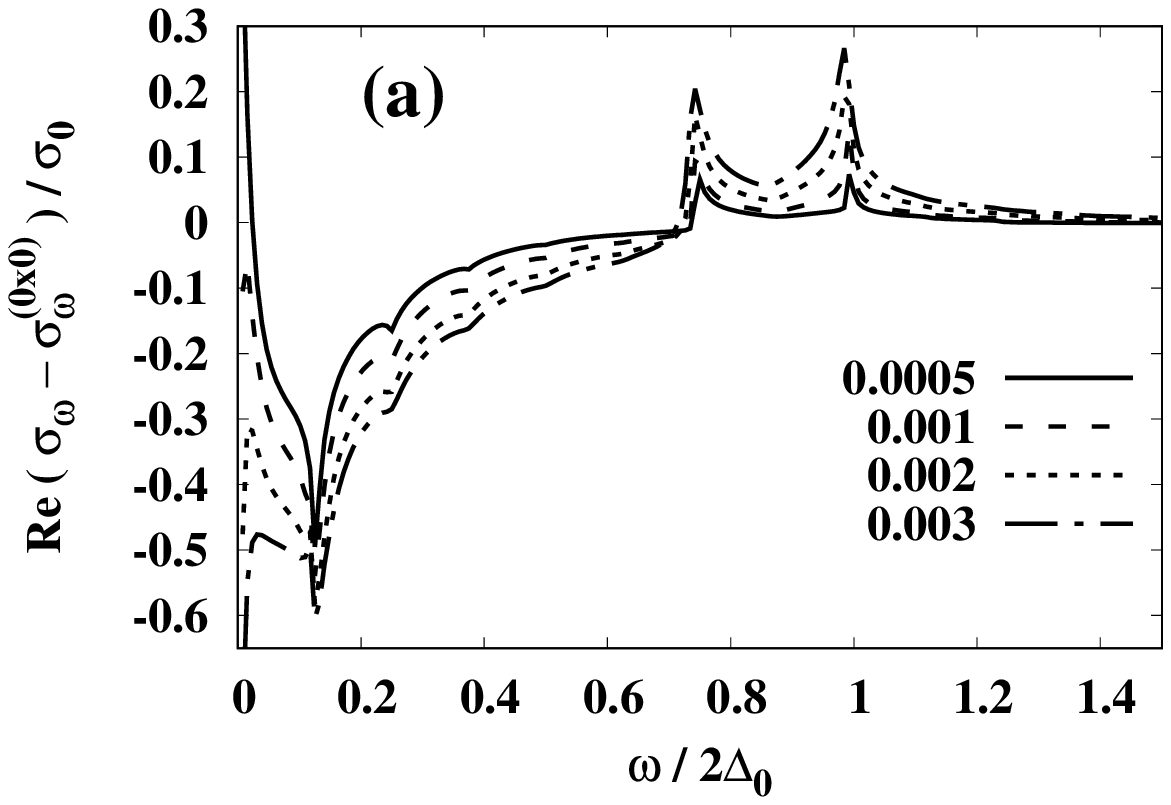}
  \includegraphics[width=8.5cm]{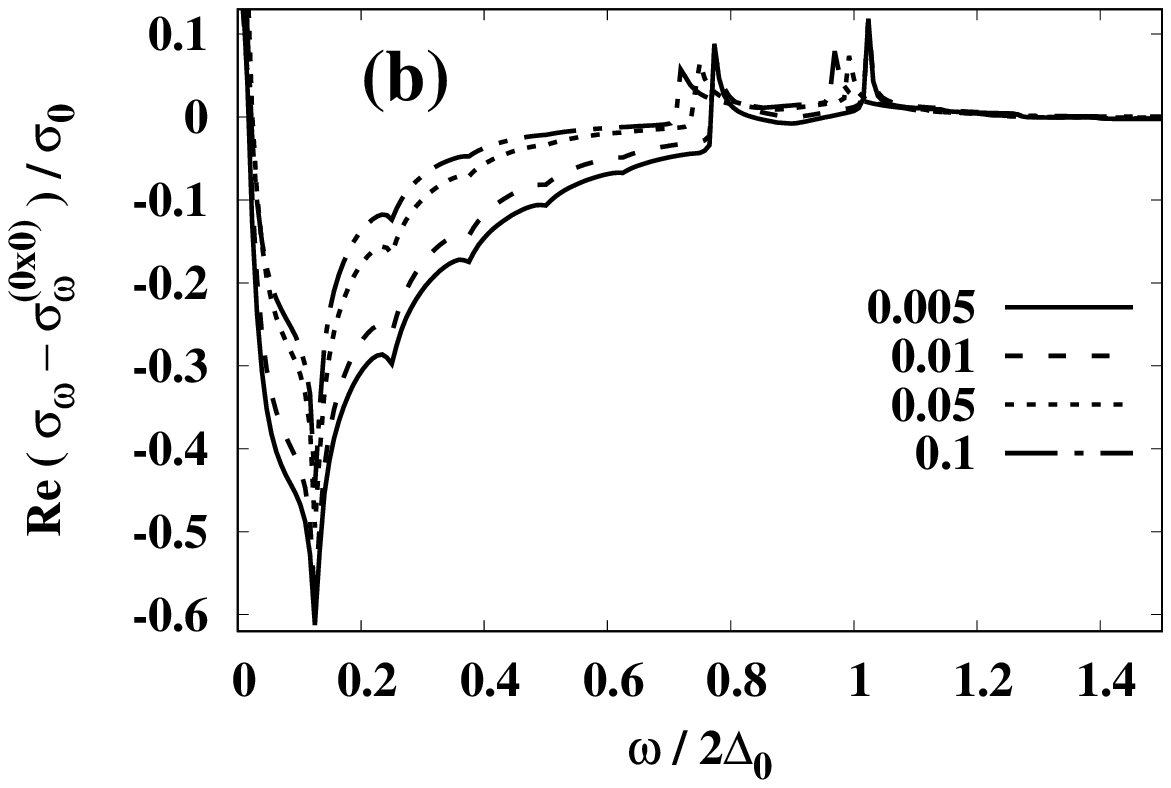}
  \caption{
    \label{fig:6}
    (a)    Dependences of ${\rm Re}(\sigma_{\omega}-\sigma^{(0x0)}_{\omega})$
    on $\omega$ for
    $\alpha/\Delta_0=0.0005$, $0.001$, $0.002$, and $0.003$.
    $p'/\Delta_0=0.05$.
    (b)    Dependences of ${\rm Re}(\sigma_{\omega}-\sigma^{(0x0)}_{\omega})$
    on $\omega$ for
    $p'/\Delta_0=0.005$, $0.01$, $0.05$, and $0.1$.
    $\alpha/\Delta_0=0.0005$.
    ($T/T_c=0.7$ and $\omega_0=0.25$.)
  }
\end{figure}

The dependences of
the nonequilibrium contribution to the absorption spectrum on
$\alpha$ and $p'$ are shown in
Figs.~\ref{fig:6}(a) and 6(b), respectively.
Figure 6(a) shows that 
the suppression of the absorption spectrum for
$\omega<2\Delta$ becomes large
and the peaks of the amplitude mode are enhanced 
as $\alpha$ increases.
This is because both 
the nonequilibrium distribution function $x^{(a)}_{\epsilon}$
and the amplitude mode $\Sigma^{\pm 1}_{a,\omega_{\pm 1}}$
are roughly proportional to $\alpha$,
which can be seen from Eqs. (\ref{eq:eqforxa}),
(\ref{eq:ampsigma}), (\ref{eq:condvc}), (\ref{eq:eqforyspm}),
and (\ref{eq:eqforzsa}).
In the same way,
the small inelastic scattering leads to 
the suppression of the absorption spectrum
for $\omega<2\Delta$ because of 
the enhancement of $x^{(a)}_{\epsilon}$.
[The denominator of Eq. (\ref{eq:eqforxa}) contains
  the inelastic scattering
  through Eqs. (\ref{eq:deftileps}) and (\ref{eq:deftildel}),
  as shown below.]
The broadening of peaks of the amplitude mode in Fig. 6(b)
originates from the damping effect.
From Eq. (\ref{eq:defsigacphpm}),
$p'=0.05\Delta_0$
gives
$|{\rm Im}\Sigma^+_{n,\epsilon=\Delta_0}|
\simeq 0.0038\Delta_0$
at $T=\Delta_0$ in the normal state,
and this gives 
an inelastic scattering time of $2.2$ ns 
in the case of $\Delta_0=0.5$ meV.
The lifetime due to the inelastic scattering
is considered to be on the order of
nanoseconds,~\cite{kaplan}
although experimental values depend on
the material.

\begin{figure}
  \includegraphics[width=8.5cm]{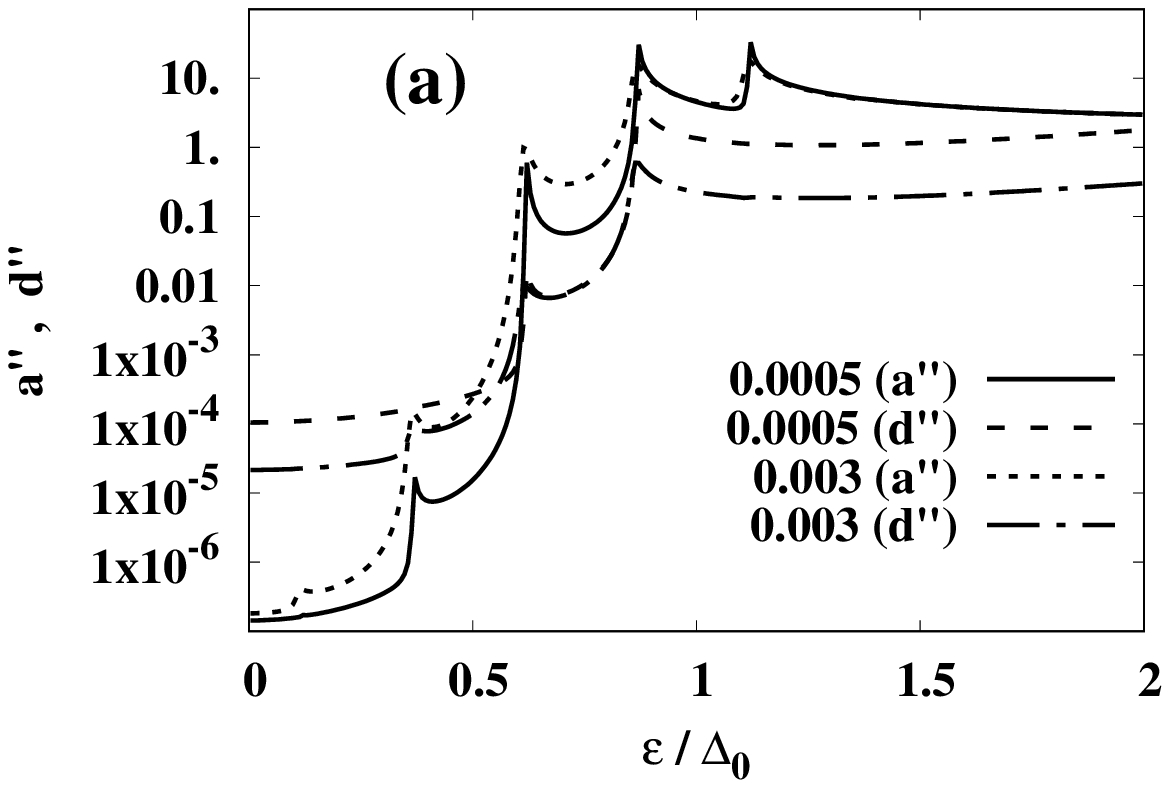}
  \includegraphics[width=8.5cm]{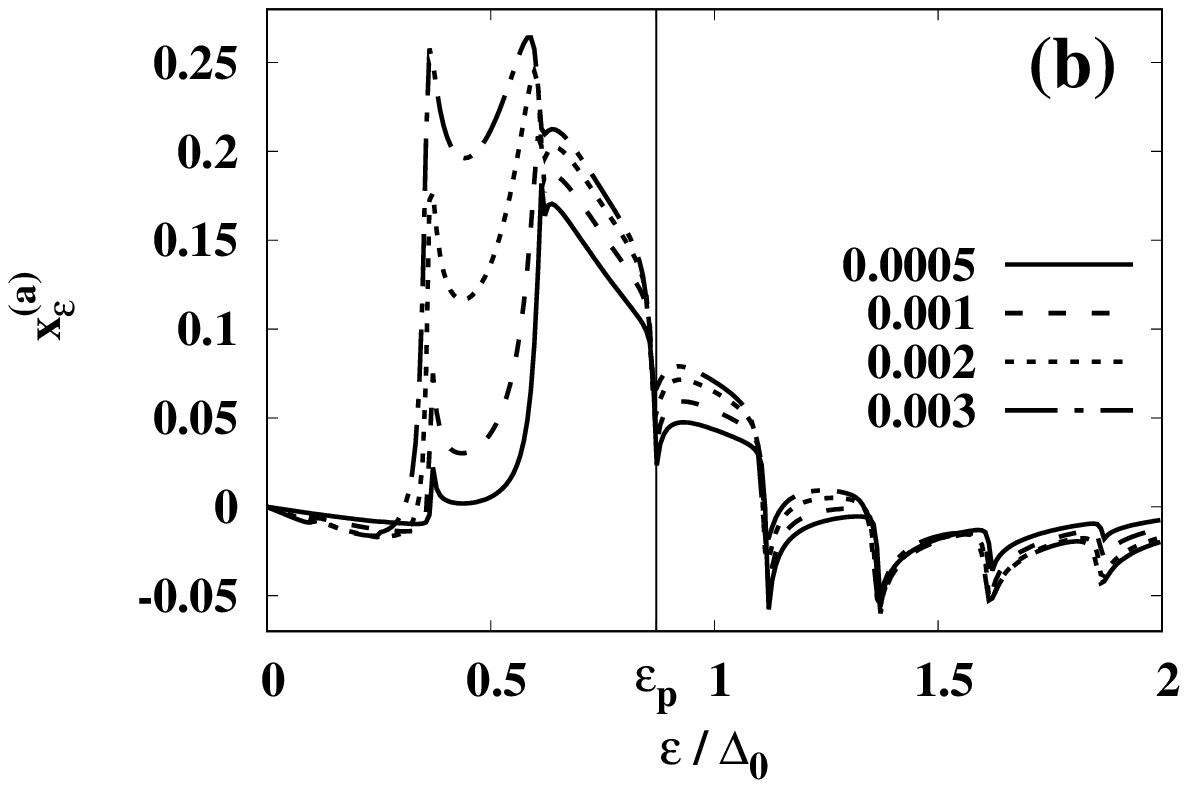}
  \caption{
    \label{fig:7}
    (a) Dependences of $a'':=(a_0+a_{-1})/\alpha$ and
    $d'':=d_0^{\gamma}/\alpha$ on $\epsilon$ for 
    $\alpha/\Delta_0=0.0005$ and $0.003$.
    (b) Dependences of $x^{(a)}_{\epsilon}$ on $\epsilon$ 
    for $\alpha/\Delta_0=0.0005$, $0.001$, $0.002$, and $0.003$.
    $T/T_c=0.7$, $\omega_0/\Delta_0=0.25$, and $p'/\Delta_0=0.05$.
    ($\epsilon_p:=\Delta/\Delta_0$.)
  }
\end{figure}

Here, we consider
the relationship between 
$\alpha$ and the inelastic scattering
in order to 
understand the absorption spectrum
around $\omega\simeq 0$ in the nonequilibrium state.
With the use of Eqs. (\ref{eq:deftileps}), (\ref{eq:deftildel}),
(\ref{eq:defaj}), and (\ref{eq:defdjgam}), 
the denominator of $x^{(a)}_{\epsilon}$ in Eq. (\ref{eq:eqforxa})
is written as 
\begin{equation}
          {\left(\tilde{\epsilon}^+-\tilde{\epsilon}^-\right)
\left(g^+_{\epsilon}-g^-_{\epsilon}\right)
    -(\tilde{\Delta}^+_{\epsilon} -\tilde{\Delta}^-_{\epsilon})
    \left(f^+_{\epsilon}-f^-_{\epsilon}\right)}
          =
a_0+a_{-1}+d_0^{\gamma}
\end{equation}
with
\begin{equation}
  (a_0+a_{-1})/\alpha=
  4\sum_{\pm}\left(
  {\rm Im}g^+_{\epsilon}{\rm Im}g^+_{\epsilon\pm\omega_0}
  +  {\rm Im}f^+_{\epsilon}{\rm Im}f^+_{\epsilon\pm\omega_0}\right).
  \end{equation}
The calculated results of 
$a'':=(a_0+a_{-1})/\alpha$ and $d'':=d_0^{\gamma}/\alpha$
are shown in Fig.~\ref{fig:7}(a). 
In the case of large $\alpha$ $(=0.003)$,
the term of coupling densities of states by microwave irradiation ($a''$)
becomes large compared with the damping effect ($d''$)
in the range of $\Delta-2\omega_0<\epsilon<\Delta-\omega_0$.
This enhances $x^{(a)}_{\epsilon}$
in this range, as shown in Fig. 7(b).
[In the case of small $\alpha$ $(=0.00005)$,
$x^{(a)}_{\epsilon}$ is small in this range of $\epsilon$
because $a''<d''$.]
The $\epsilon$ dependence of
$x^{(a)}_{\epsilon}$ for $\epsilon>\Delta$
is also varied by 
this enhancement of $x^{(a)}_{\epsilon}$
through the coupling between $x^{(a)}_{\epsilon}$
and $x^{(a)}_{\epsilon\pm\omega_0}$,
as shown in Eq. (\ref{eq:eqforxa}).
($\epsilon_p=\Delta/\Delta_0$ indicates that
the density of states 
takes large values for $\epsilon/\Delta_0\ge\epsilon_p$.)
Equation (\ref{eq:resig0bytau})
indicates that the absorption spectrum around $\omega\simeq 0$
is determined by the values of
$dx^{(a)}_{\epsilon}/d\epsilon$
for $\epsilon/\Delta_0\ge\epsilon_p$. 
The enhancement of $x^{(a)}_{\epsilon}$ mentioned above
makes
negative values of $dx^{(a)}_{\epsilon}/d\epsilon$
predominant over
positive values of $dx^{(a)}_{\epsilon}/d\epsilon$
in the absorption spectrum
on increasing the value of $\alpha$.
This explains the variation of the spectrum 
around $\omega\simeq 0$ in Fig. 6(a),
and the same tendency holds when
the value of $p'$ is varied, although
the effect seems small in Fig. 6(b).

\section{Summary and Discussion}

In this paper, we calculated the absorption spectrum of
$s$-wave superconductors in the nonequilibrium steady state
under microwave irradiation.
In this state, the distribution function of
quasiparticles is
reduced on the low-energy side and shifted to
the high-energy side compared with that in the equilibrium
state.
This nonequilibrium redistribution
is large around energies across the superconducting gap,
so it is important to include the
distribution function at energies
not only above the superconducting gap
but also below this gap
in the calculation of the absorption spectrum.

Previously, the distribution function
was calculated
only for energies above the superconducting gap
and a reduction of 20 or 25 percent was obtained in
the absorption spectrum.~\cite{chang78,sridhar}
In this study,
the calculated reduction of the absorption is 
about 59 percent (the value of
$|{\rm Re}\sigma_{\omega_0}^{(0)}-{\rm Re}\sigma_{\omega_0}^{(0x0)}|/
{\rm Re}\sigma_{\omega_0}^{(0x0)}$ at $T/T_c=0.7$ for
$\alpha/\Delta_0=0.0005$).
Therefore, most of the decrease in the absorption spectrum owing
to the nonequilibrium state is
attributed to the changes in the distribution function
at energies below the superconducting gap.

Since the energy dependence of the nonequilibrium
distribution function
cannot be described by an exponential function
with a single temperature,
there is no single parameter that
characterizes the nonequilibrium state.
The absorption spectrum 
reflects this nonuniform function of energy,
and thus,
it is not valid to relate the magnitude of
the absorption and the effective temperature
in the nonequilibrium state as was done in Ref. 34.
Actually, as shown in Fig. 2,
the nonequilibrium distribution
is seemingly approximated
by ${\rm tanh}(\epsilon/2T)$ 
when it is averaged over an energy scale of
several times of $\omega_0$.
The reduction of the absorption spectrum in
the nonequilibrium state is caused by
the variation over this energy scale of $\omega_0$
in the distribution function.

In the nonequilibrium state, peaks appear in
the absorption spectrum because
the amplitude modes couple with the electromagnetic field.
Since the nonequilibrium state is due to
the irradiation of the microwave,
there are two peaks on both the high-frequency 
and the low-frequency sides of the absorption
edge of the equilibrium-state spectrum.
This is in contrast to
the single peak under a static magnetic field.

In previous studies, the spectrum
originating from the amplitude mode was
calculated perturbatively with respect to
the magnetic field or microwave intensities.~\cite{jujo17,moor}
This study showed that the peak in the spectrum persists even if
the calculation is 
performed with this intensity taken into account
nonperturbatively.
$\alpha$ defined in Eq. (\ref{eq:defalpha})
is rewritten as
$\alpha/\Delta_0=(\Delta_0\sigma_0/\omega_0|\sigma_{\omega_0}|)
\left(|H_{\omega_0}|/H_c\right)^2$
with $H_c$ and $H_{\omega_0}$ being
the thermodynamic critical field at absolute zero
and the rf magnetic field,
respectively.
$|H_{\omega_0}|/H_c=\sqrt{
\left(\omega_0|\sigma_{\omega_0}|/\Delta_0\sigma_0\right)
(\alpha/\Delta_0)}\simeq 0.088$ at $T/T_c=0.7$ for
$\alpha/\Delta_0=0.003$
when using a numerically calculated result of
the absolute value $|\sigma_{\omega_0}/\sigma_0|$.
Our results show that 
the amplitude mode can be observed
at low microwave intensities below the critical
field, and this is possible to verify by experimental studies.

In this study, the polarization direction of
the probe wave is assumed
to be parallel to that of the pump wave,
but if these directions are taken to be orthogonal
to each other,
the vertex correction term is not probed in the absorption
spectrum, and there is no peak originating from the amplitude mode.
On the other hand,
the reduction in the absorption spectrum
is caused by 
the redistribution of quasiparticles,
which is irrelevant to the vertex correction.
The dependence of this reduction on
the polarization direction is small,
which should also be observed in experiments.

\section*{Acknowledgment}

The numerical computation in this work was carried out
at the Yukawa Institute Computer Facility.


\begin{thebibliography}{9}
\bibitem{memo1}   This is the result in the case of $\alpha\ne 0$.
  In the absence of the pump wave ($\alpha= 0$),
  there is no vertex correction, and $x^{(a)}_{\epsilon}=0$
  from Eq. (\ref{eq:eqforxa}). In this case, the
  distribution function remains the same as that in
  the equilibrium state, which leads to
  the absorption spectrum for the linear response.
\end{thebibliography}
\end{document}